\documentclass[default, pdflatex]{sn-jnl}% Default
\usepackage{amssymb}%
\usepackage{graphicx}%
\renewenvironment{table}[1][]%
{\tableorg[#1]%
\tablebodyfont%
\renewcommand\footnotetext[2][]{{\removelastskip\vskip3pt%
\let\tablebodyfont\tablefootnotefont%
\hskip0pt\if!##1!\else{\smash{$^{##1}$}}\fi##2\par}}%
}{\endtableorg}
\usepackage{float}

\usepackage{xcolor}%
\usepackage{booktabs}\let\cline\cmidrule
\usepackage{multicol}
\usepackage{multirow}%
\usepackage{subcaption}
\usepackage{longtable}
\usepackage{enumitem}
\usepackage{hyperref}

\usepackage{manyfoot}
\usepackage{amsmath}

\usepackage{xspace}
\newcommand{\pt}           {\ensuremath{p_{\rm T}}\xspace}

\newcommand{\nineH}        {$\sqrt{s}=0.9$~Te\kern-.1emV\xspace}
\newcommand{\seven}        {$\sqrt{s}=7$~Te\kern-.1emV\xspace}
\newcommand{\twoH}         {$\sqrt{s}=0.2$~Te\kern-.1emV\xspace}
\newcommand{\twosevensix}  {$\sqrt{s}=2.76$~Te\kern-.1emV\xspace}
\newcommand{\five}         {$\sqrt{s}=5.02$~Te\kern-.1emV\xspace}
\newcommand{\thirteen}     {$\sqrt{s}=13$~Te\kern-.1emV\xspace}
\newcommand{\twosevensixnn}{$\sqrt{s_{\mathrm{NN}}}=2.76$~Te\kern-.1emV\xspace}
\newcommand{\fivenn}       {$\sqrt{s_{\mathrm{NN}}}=5.02$~Te\kern-.1emV\xspace}

\newcommand{\TeV}          {Te\kern-.1emV\xspace}
\newcommand{\GeV}          {Ge\kern-.1emV\xspace}
\newcommand{\MeV}          {Me\kern-.1emV\xspace}
\newcommand{\GeVmass}      {\ensuremath{\rm{Ge\kern-.1emV/}c^2\xspace}}
\newcommand{\MeVmass}      {\ensuremath{\rm{Me\kern-.1emV/}c^2\xspace}}
\newcommand{\GeVc}         {\ensuremath{\rm{Ge\kern-.1emV/}c\xspace}}
\newcommand{\MeVc}         {\ensuremath{\rm{Me\kern-.1emV/}c\xspace}}
\newcommand{\eVc}         {\ensuremath{\rm{e\kern-.1emV/}c\xspace}}
\newcommand{\GeVcc}        {\ensuremath{\rm{Ge\kern-.1emV/}c^2\xspace}}
\newcommand{\MeVcc}        {\ensuremath{\rm{Me\kern-.1emV/}c^2\xspace}}
\newcommand{\eVcc}        {\ensuremath{\rm{e\kern-.1emV/}c^2\xspace}}

\let\GeVmass=\GeVcc
\let\MeVmass=\MeVcc

\begin{document}

\title[Article Title]{Machine-learning-based particle identification with missing data}


\author[a]{\fnm{Miłosz} \sur{Kasak}}
\author[a,b]{\fnm{Kamil} \sur{Deja}}%\corref{author}}
\author[a,c]{\fnm{Maja} \sur{Karwowska}}
\author[a]{\fnm{Monika} \sur{Jakubowska}}
\author[a]{\fnm{Łukasz} \sur{Graczykowski}}
\author[a]{\fnm{Małgorzata} \sur{Janik}}

% \cortext[author] {Corresponding author. \textit{E-mail address:} kamil.deja@pw.edu.pl}
% \affil[a]{\orgdiv{Faculty of Electronics and Information Technology, Warsaw University of Technology}, \orgaddress{\street{Nowowiejska 15/19}, \postcode{00-665} , \city{Warsaw}, \country{Poland}}}
% \affil[b]{\orgdiv{Faculty of Physics, Warsaw University of Technology} \orgaddress{\street{Koszykowa 75}, \postcode{00-662}, \city{Warsaw}, \country{Poland}}}
% \affil[c]{{\orgdiv{Faculty of Electrical Engineering, Warsaw University of Technology} \orgaddress{\street{Koszykowa 75}, \postcode{00-662} , \city{Warsaw}, \country{Poland}}}}

\affil[a]{{\orgdiv{Warsaw University of Technology}, \orgaddress{\street{pl. Politechniki 1}, \postcode{00-661} , \city{Warsaw}, \country{Poland}}}}
\affil[b]{\orgdiv{IDEAS NCBR},  \orgaddress{\street{Chmielna 69}, \postcode{00-801}, \city{Warsaw}, \country{Poland}}}
\affil[c]{\orgdiv{CERN -- European Organization for Nuclear Research},  \orgaddress{\street{Espl. des Particules 1}, \postcode{1211} \city{Geneva}, \country{Switzerland}}}

% \author*[1,2]{\fnm{First} \sur{Author}}\email{iauthor@gmail.com}



% \affil*[1]{\orgdiv{Department}, \orgname{Organization}, \orgaddress{\street{Street}, \city{City}, \postcode{100190}, \state{State}, \country{Country}}}

% \affil[2]{\orgdiv{Department}, \orgname{Organization}, \orgaddress{\street{Street}, \city{City}, \postcode{10587}, \state{State}, \country{Country}}}

% \affil[3]{\orgdiv{Department}, \orgname{Organization}, \orgaddress{\street{Street}, \city{City}, \postcode{610101}, \state{State}, \country{Country}}}


\abstract{In this work, we introduce a novel method for Particle Identification (PID) within the scope of the ALICE experiment at the Large Hadron Collider at CERN. Identifying products of ultrarelativisitc collisions delivered by the LHC is one of the crucial objectives of ALICE. Typically employed PID methods rely on hand-crafted selections, which compare experimental data to theoretical simulations. To improve the performance of the baseline methods, novel approaches use machine learning models that learn the proper assignment in a classification task. However, because of the various detection techniques used by different subdetectors, as well as the limited detector efficiency and acceptance, produced particles do not always yield signals in all of the ALICE components. This results in data with missing values. Out of the box machine learning solutions cannot be trained with such examples without either modifying the training dataset or re-designing the model architecture.
% , so a significant part of the data is skipped during training. 
In this work, we propose the new method for PID that addresses these issues and can be trained with all of the available data examples, including incomplete ones. Our approach improves the PID purity and efficiency of the selected sample for all investigated particle species.
}




\maketitle

\section{Introduction}
ALICE (A Large Ion Collider Experiment)~\cite{ALICE:2008ngc} is one of the four major detectors located at the Large Hadron Collider at CERN~\cite{Evans:2008zzb}. The main goal of ALICE is to study the properties of quark--gluon plasma (QGP), a hot and dense state of matter, and the strong force that binds quarks together inside hadrons~\cite{ALICE:2022wpn}. The key requirement for detailed studies of QGP that distinguishes ALICE from the other  Large Hadron Collider (LHC) experiments is its capability for very precise particle identification (PID) -- i.e. the ability to discriminate between different particle species produced during the collision. This allows for selecting a subset of particles required for specific analysis.

\begin{figure}
    \begin{center}
    \includegraphics[width = .85\textwidth]{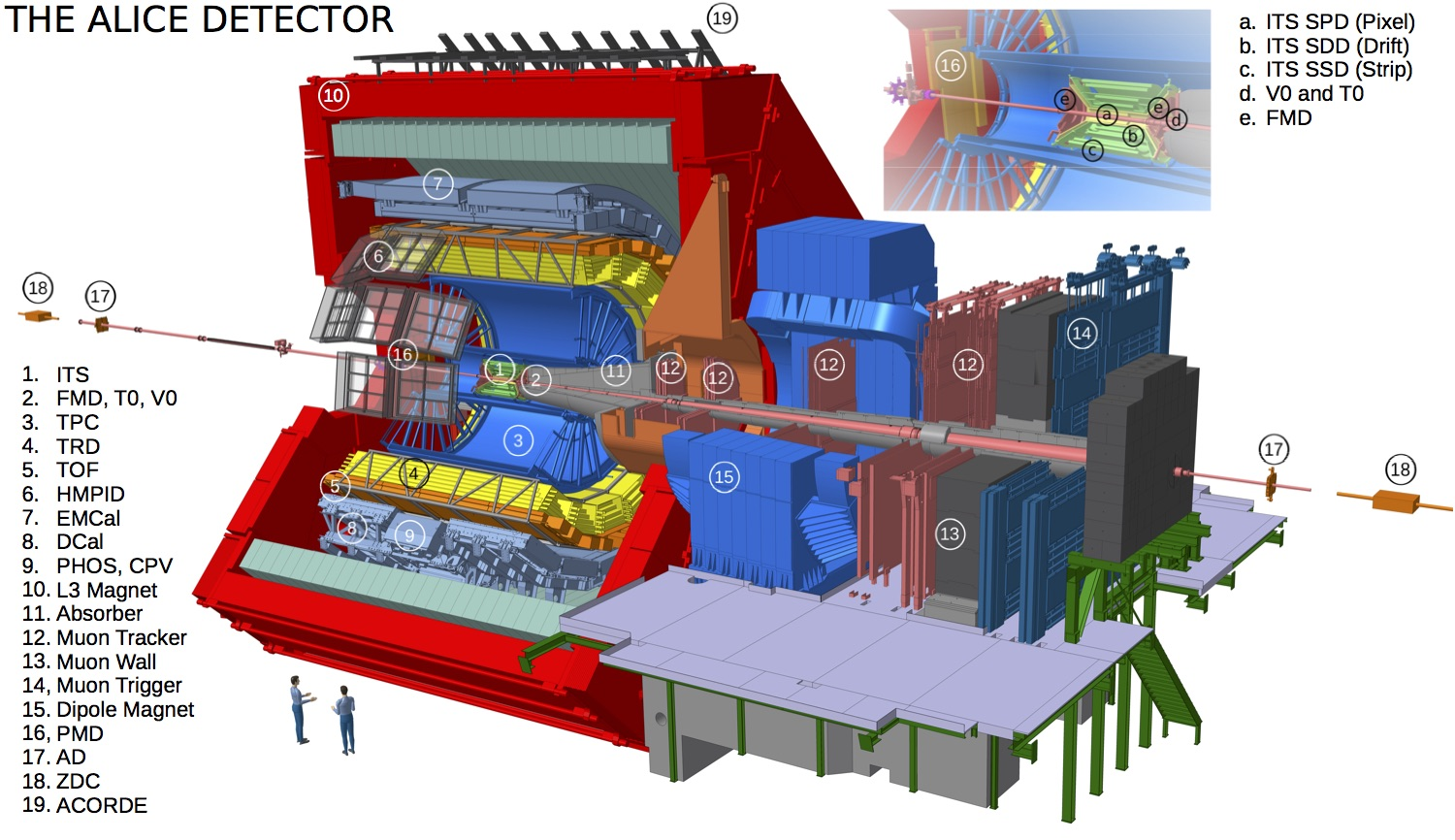}
    \end{center}
    \caption{Components of the~ALICE detector in its Run 2 configuration \cite{botta:2017a}.}
    \label{fig:alice-detectors}
\end{figure}

The ALICE experiment is composed of several sub-detectors, some of which measure particle properties that can be used for identification. Figure \ref{fig:alice-detectors} presents a scheme of the detector in Run 2, the previous LHC data-taking periods.

The three detectors particularly useful for PID are: TPC, TOF, and TRD. The Time Projection Chamber~\cite{tpc-tdr} (TPC) is one of the most important ALICE detectors as it records 3D information of the~trajectory of charged particles, as well as their specific energy loss due to ionization, which is essential for particle identification. The Time-of-Flight~\cite{tof-tdr} (TOF) detector measures  particle travel times from the collision vertex to the detector, from which the particle velocity and mass are calculated. The Transition Radiation Detector~\cite{trd-tdr} (TRD) records transition radiation, that is, the emission of photons by electrons traversing the boundaries of a radiator, which helps in distinguishing electrons from other charged particles. All the detectors mentioned above detect particles carrying a non-zero electric charge. Therefore, this article focuses on the identification of charged particles.

With the signals recorded by the detectors described above, particles are chosen using a set of selection criteria.
%calculated using a simulation of the particle transport through the detector geometry with GEANT~\cite{Brun:1994aa} system followed by a detailed detector response simulation.
Traditionally, the particle identification is based on hand-crafted selection criteria, for instance, based on how much the detector response deviates from the expected value. Particles falling outside the selection region are rejected. One of the most common selection methods is described in detail in Section~\ref{sec:pid}.

However, when the characteristics of different particle species overlap, combining information from multiple detectors becomes difficult. Choosing selection regions by trial and error is less effective in this case, lowering the purity of the selected sample.

These shortcomings can be addressed with more advanced classification methods such as Bayesian models~\cite{ALICE:2016zzl} or neural-network-based (NN) approaches~\cite{lhcb2015lhcb, collado2021learning, cms2022identification}. 
However, the more sophisticated methods can only be trained or employed with complete data instances; just as a function value cannot be computed without all the inputs, NNs may not be used to process examples with missing values.
As detectors used for particle identification work independently, particles may be measured by a subset of detectors while not being recorded by others. This can happen for various reasons; for instance, a detector may malfunction or be switched off, or the particle can have characteristics (e.g. too low transverse momentum) that do not match the detector specification. Collected data may therefore contain incomplete measurements, with information from only some detectors.
As a result, standard neural-network-based approaches cannot be used for PID without altering either the incomplete data or the model architecture. 
The former approach is usually implemented by inputting average or predicted missing values. This introduces artificial information to the data that can change the predictions of a model in undefined ways. Such an approach can often lead to biased outputs, which is especially problematic for real-life applications such as PID, where model predictions may lead to physically impossible outcomes.
Therefore, we focus on altering the model architecture, as this approach allows us to avoid making any assumptions about the missing data and ensures that no contained information is modified.

In this work, we introduce the new method of PID with missing data. In particular, we propose an adaptation of an attention-based multi-instance learning approach ~\cite{wang2019attention}. The resulting model can predict and learn from examples containing any number of missing measurements and allows for using knowledge learned from complete examples when classifying incomplete ones, as well as the other way around.

Our tests show that this additional knowledge improves the performance of neural-network-based models. We evaluate different methods for coping with incomplete data and show that our approach achieves state-of-the-art performance in PID with neural networks. We attribute this performance gain to the proposed method that learns from all available data, including examples with missing data. % currently used PID methods as well as other state-of-the-art approaches.
% %
\section{Related works}

\subsection{Missing data sources}
The incomplete data problem in statistical and machine learning (ML) models has been widely studied over decades~\cite{garcia2010pattern,ghahramani1995learning}. In~\cite{LittleRoderickJ.A2014Sawm}, authors highlight three main categories of missing data according to the source of the information gaps. The first case occurs when examples are \textit{missing completely at random} (MCAR) -- where the probability of missing value is independent of all parameters. If the probability of the missing value is dependent only on the value of the observed attributes, we can call it \textit{missing at random} (MAR). Finally, if this probability depends on the value of the missing parameter itself, we refer to this case as \textit{missing not at random} (MNAR).

In our case of the data used for particle identification at the ALICE experiment, we can distinguish two main sources of the missing data. For some cases, the lack of particular value -- signal from one of the detectors is caused by its temporal malfunction, independent of the measured particle. In this case the signal is missing completely at random (MCAR). On the other hand, for some cases we might lack the measurement of values that fall outside the effective region of particular detector. For example, the efficiency of the TOF detector measurements drop significantly for particles with transverse momentum below 0.5 \GeVc. In such case, the missing data is missing not at random (MNAR).

\subsection{Machine learning with incomplete data}

Several popular machine learning algorithms such as neural networks, cannot be directly trained with missing data. Therefeore there exists different solutions tot that problem that can be roughly divided into two groups: (1) those that modify the training dataset~\cite{jadhav2019comparison,young2011survey} and (2) those that adapt ML architectures~\cite{jiang2005classification,juszczak2004combining,krause2003ensemble,sharpe1995dealing}.
% Most focus is put on treating the missing values within datasets~\cite{jadhav2019comparison,young2011survey}, as well as adapted ML architectures, including neural network ensembles~\cite{jiang2005classification,juszczak2004combining,krause2003ensemble,sharpe1995dealing}.
Data treatment methods transform datasets with missing values into complete ones that can be used with standard machine learning architectures.
Adapted architectures are capable of processing incomplete data without any alteration.

The simplest method for data transformation is known as \emph{case deletion}, where examples without full data availability are simply removed from the training set. This limits the potential training capabilities of the model and can result in bias induced by missing part of the original data distribution. The alternative technique known as \emph{imputation} fills the missing data with artificial values calculated as a \emph{mean}, \emph{median}, or value predicted with a simple model, e.g. \emph{linear regression}. This idea is further extended in~\cite{morvan2021s}, where the imputation model is optimized jointly with the regressor. Similarly to the case deletion approach, imputation methods can also significantly disturb the predictions of the neural model, which might result in unrealistic behavior. 

To overcome those shortcomings, methods based on the model's adjustment were proposed. The most straightforward approach suitable in a situation when missing data affects only a few attributes is an ensemble of different classifiers. 
In such a case, different models are trained on subsets of the training dataset without missing data. 
In particular, in \emph{neural network reduction}~\cite{sharpe1995dealing}, authors propose to split the dataset into the largest possible complete subsets. 

There are several domains where machine learning with incomplete data is already employed. In~\cite{lipton2016modeling}, authors use a recurrent neural network to model healthcare data with missing data. A similar idea was employed in medical applications such as breast cancer prediction~\cite{jerez2010missing}. Machine learning algorithms with missing data are also used in other domains such as traffic prediction~\cite{shi2015improving,cui2020stacked}.

\subsection{Particle identification techniques}
\label{sec:pid}
As mentioned in the introduction, the recently employed PID techniques depend on the human-defined selection criteria on the response signal from given detectors used in the analysis. In most cases, the so-called ``$\rm n_{\sigma}$'' method is used, whose main parameter is the number of standard deviations from the expected value that a signal for a given particle left in the detector. For example, if both the TPC and TOF detectors are used, a common approach is to define a PID selection as 
\begin{equation}
    \rm \sqrt{n_{\sigma,TPC}^{2}+n_{\sigma,TOF}^{2}}<\lambda.
\end{equation}

The actual cut-off value $\rm \lambda$, however, depends on the specific analysis, and requirements for purity and efficiency of the sample. Typically, this value is set between 2 and 3. This method can be further modified by adding additional conditions, for instance, on rejection of other types of particles. In such a case, if one analyzes, i.e., charged kaons, one may provide the acceptance criterion for the kaon hypothesis (i.e. how far in terms of $\rm n_{\sigma}$ the signal can be from the expected value for kaons) as well as the rejection criterion for the pion, proton, and electron hypothesis (i.e. what is the minimum $\rm n_{\sigma}$ that the signal must have compared to the expected values for pions, protons, and electrons).

% The other, less commonly used PID technique is based on the Bayesian approach~\cite{ALICE:2016zzl}. This method requires the definition of priors, which serve as a ``best guess'' of true particle yields per event. The main motivation is the analysis of rare decays with high levels of combinatorial background like \LctopKpi whose identification with the $\rm n_{\sigma}$ method would not be possible in the previous data-taking period. The Bayesian PID was tested in the case of identification of \DtoKpi decay, yielding a higher signal-to-background ratio than the standard $\rm n_{\sigma}$ approach.

Outside of the ALICE experiment, several attempts have been made to use neural networks for the PID. In the LHCb experiment~\cite{lhcb2015lhcb}, shallow neural networks are used to classify calorimeter signals. Similarly, in ATLAS, neural models are used to identify electrons~\cite{collado2021learning}, while the CMS Collaboration identifies $\tau$ lepton decays with deep neural networks~\cite{cms2022identification}.

% \section{The Particle Identification as classification task with missing data}

\section{Attention-based neural network architecture for particle identification}

\begin{figure}[t]
\includegraphics[width=\textwidth]{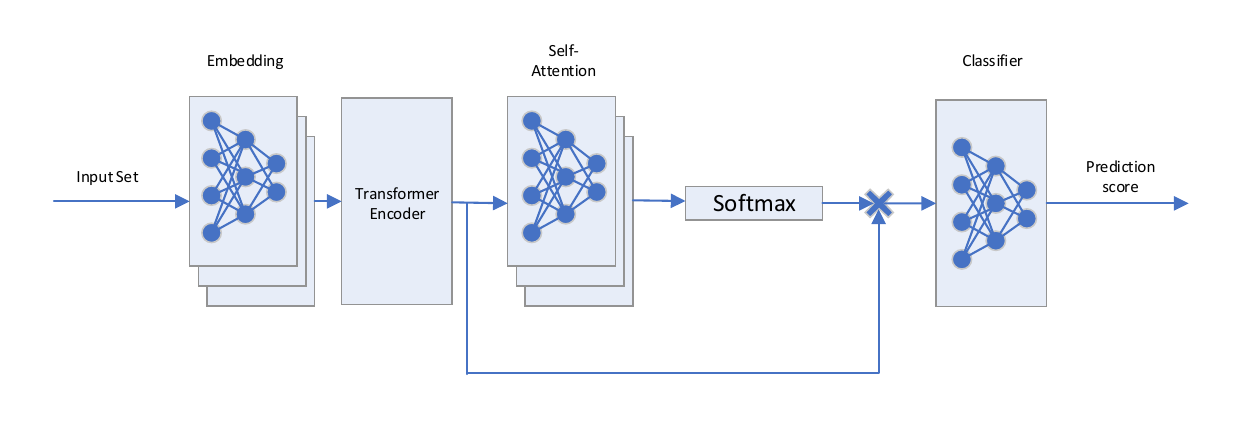}
\caption{The proposed model architecture. Layered blocks are applied separately to each vector in a set. Single blocks are applied to their input as a whole.} \label{fig:model_arch}
\end{figure}

In this work, we focus on the problem of PID in which a type of particle is assigned to a data sample based on its characteristics recorded by a set of detectors.
We treat particle identification as a set of binary classification tasks in a so-called \emph{one vs.\ all} approach.
Given a set of examples, each labeled with one of \(k\) particle types, the task is to find \(k\) binary classifiers corresponding to each particle type.
Each classifier is tasked with identifying whether or not a particle, represented by a given example, belongs to the specific particle type.
Every example consists of a set of real-valued features, based on measurements from a detector, out of which some might be missing. 

With this formulation, we introduce a novel method for PID that employs missing data. We base our approach on the attention mechanism, similar to the method introduced for a medical use-case in AMI-Net~\cite{wang2019attention}. Our system is composed of several steps.
We first encode all data examples into a set of feature-value pairs~\cite{grangier2010feature}, independently of their characteristics regarding missing data, and transform them into embeddings. These embeddings are further processed by the Transformer's~\cite{vaswani2017attention} encoder module with multi-head attention. The encoder output is combined into a single feature vector that is an input to the final classifier.
The overview of our system is shown in Fig. \ref{fig:model_arch}, while in the following subsections, we describe its building blocks in detail.

\subsection{Feature Value Pairs}
We prepare the incomplete data for the model by creating sets of feature-value pairs from the incomplete vectors.
Each pair consists of a non-missing value in the vectors and the index of that value.
We apply one-hot encoding to these indices to aid the model in their processing.
An example is shown in table \ref{fse_example}.

\begin{table}
    \renewcommand{\arraystretch}{1.2}
    \setlength{\tabcolsep}{5pt}
    \footnotesize
    \centering
    \caption{An example for preprocessing of data samples into feature set values.}
    \label{fse_example}
    % \resizebox{\textwidth}{!}{ 
        \begin{subtable}[t][][b]{\textwidth}
            \caption{3 data samples with 4 attributes with different amounts of missing values.}
            \centering
            \begin{tabular}{|l|r|r|r|r|}
                \hline
                id  & momentum &TOF &  TPC & TRD \\
                \hline
                1 & 0.1 &  & 3 &  \\
                2 & 7 & 70 & 24 & 13 \\
                3 &  & 78 &  &  \\
                \hline
            \end{tabular}
        \end{subtable}
        \hfill
        \vskip 1pt
        %\vspace{1cm}
        \begin{subtable}[t]{0.25\textwidth}
            \caption{First particle.}
            \begin{tabular}{|c|c|c|c|r|}
                \hline
                \multicolumn{4}{|c|}{key} & value \\
                \hline
                1 & 0 & 0 & 0 & 0.1\\
                0 & 0 & 1 & 0 & 3\\
                % 0 & 0 & 0 & 0 & 1 & 5\\
                & & & & \\
                & & & & \\
                \hline
            \end{tabular}
        \end{subtable}
        \hspace{1cm}
        % \hfill
        \begin{subtable}[t][][b]{0.25\textwidth}
        \caption{Second particle.}
            \begin{tabular}{|c|c|c|c|r|}
                \hline
                \multicolumn{4}{|c|}{key} & value \\
                \hline
                1 & 0 & 0 & 0  & 7\\
                0 & 1 & 0 & 0  & 70\\
                0 & 0 & 1 & 0  & 24\\
                0 & 0 & 0 & 1  & 13\\
                % 0 & 0 & 0 & 0 & 1 & 88\\
                \hline
            \end{tabular}
        \end{subtable}
        % \hfill
        \hspace{1cm}
        \begin{subtable}[t][][b]{0.25\textwidth}
            \caption{Third particle.}
            \begin{tabular}{|c|c|c|c|r|}
                \hline
                \multicolumn{4}{|c|}{key} & value \\
                \hline
                0 & 1 & 0 & 0 & 78\\
                 & & & &  \\
                 & & & &  \\
                 & & & &  \\
                 % & & & & & \\
                \hline
            \end{tabular}
        \end{subtable}
    % }
\end{table}

\subsection{Embedding}
Embedding is a continuous vector representation of discrete data.
Embedded feature indices can use relations between features more effectively by placing similar features close in the embedded space.
We create embeddings by applying a neural network with a single hidden layer to the feature-value vectors.

\subsection{Transformer Encoder}

Transformer is an attention-based architecture commonly used for sequence transduction. As the Transformer is used to process sequences of arbitrary lengths, it can naturally be adapted to find/learn relations between available features regardless of the amount of missing values.
We apply the Transformer's encoding module to the set of embedding vectors to connect the different features each vector represents.

The encoder consists of a stack of $N$ identical layers. 
Each layer is made of two sub-layers: a multi-head attention layer and a dense neural network. The multi-head attention is applied to the input set as a whole, while the NN is applied to each vector in a set separately.
To ease the training of the encoder, we apply a residual connection around each sub-layer, followed by layer normalization.

\subsubsection{Multi-head attention}

To connect pairs of vectors in the input set, each input vector is linearly transformed into \(h\) query, key, and value vectors, and scaled dot-product attention is applied to each of the \(h\) triples of query, key, value sets, where \(h\) is the number of heads. This allows specific patterns to be found in pairs of input vectors. For example, a measurement from a specific detector could be used if the momentum is within a particular range. As each of the \(N\) layers in the encoder contains a multi-head attention sub-layer, and each sub-layer connects pairs of vectors, the full encoder can theoretically connect subsets of \(2^N\) vectors.

\subsubsection{Scaled dot-product attention}

Given sets of query, key, and value vectors of dimension \(d_k\), scaled dot-product attention calculates a set of weighted averages of the value vectors based on the similarities between the corresponding keys and queries.
The similarities are obtained by computing the dot product of the query and key vectors. The dot products are scaled by a factor of \(\frac{1}{\sqrt{d_k}}\) to counteract the increase of the dot product with an increasing dimension of the query, key, and value vectors.
Finally, the softmax function is applied to the scaled dot products to obtain the weights used to calculate the averages of value vectors. The whole computation can be described as
\begin{equation}
    \mathrm{Attention}(Q, K, V) = \mathrm{softmax}\left (\frac{QK^T}{\sqrt{d_k}}\right )V,
\end{equation}
where \(Q,K,V \in \mathbb{R}^{n \times d_k}\) are sets of \(n\) query, key and value vectors, respectively.

\subsection{Attention pooling}

As the classifier neural network cannot process an unordered, variable-size set of vectors, merging the vectors obtained from the Transformer's encoder into a single vector is necessary. 
Therefore, we pool the output vectors together by using the self-attention technique that was also used in the Transformer, where each vector is assigned weights based on its own values. These weights are then used to calculate the weighted average of all the vectors in the set.

\subsubsection {Self attention}

Given a set of vectors \(\{v_1, v_2, ..., v_n\}\), where \(v_i \in \mathbb{R}^{d_{model}}\) , we obtain the pooled vector as follows:
\begin{align}
E_{i, \star} &= \mathrm{NN}(v_i) &\forall i &\in [1,n]\\
A_{{\star}, j} &= \mathrm{softmax}(E_{{\star}, j}) &\forall j &\in [1,d_{model}]\\
o_j &= \sum_{k=1}^{n} A_{k, j}v_{k_j} &\forall j &\in [1,d_{model}]
\end{align}

Here, \(E, A \in \mathbb{R}^{n \times d_{model} }\) are respectively the self-attention values and weights, \(o \in \mathbb{R}^{d_{model}} \) is the pooled output vector, and \(NN\) is a neural network that has both dimensions of input and output equal to \(d_{model}\).

\subsection {Classifier}
We obtain the final prediction score for the PID task by applying a simple neural network with a single-value output to the pooled vector.
We normalize the score to the range \((0,1)\) by applying the logistic function:
\begin{equation}
    f(x) = \frac{1}{1+e^{-x}}.
\end{equation}
%\(f(x) = \frac{1}{1+e^{-x}}\).
The obtained score approximates the probability of the example corresponding to a specific particle type.

\section{Evaluation}
We evaluate our model on the PID task by comparing it against a standard $n_\sigma$-based selection technique, an ensemble of neural networks~\cite{sharpe1995dealing} as well as two standard techniques for ML with missing data: mean imputation, and linear regression imputation. Additionally, in a scenario with no missing data in the test-set, we compare our approach to the case deletion procedure.

\subsection{Dataset}
The data comes from a Monte Carlo simulation of proton--proton collisions at $\sqrt{s}=13$~TeV with a realistic simulation of the time evolution of the detector conditions in the LHC Run 2 data-taking period. The simulation was performed with PYTHIA 8~\cite{Sjostrand:2014zea}, the GEANT~\cite{Brun:1994aa} particle transport model, and general-purpose settings.

\begin{figure}[h!]
    \begin{minipage}[b]{.5\textwidth}
        \centering
        \resizebox{\linewidth}{!}{
            \begin{tabular}{ccccc}
\toprule
pion & kaon & proton & electron & muon \\
\midrule
43.59\% & 3.415\% & 2.026\% & 0.794\% & 0.323\% \\
\cline{1-5}
\bottomrule
\toprule
antipion & antikaon & antiproton & antielectron & antimuon \\
\midrule
43.66\% & 3.288\% & 1.819\% & 0.762\% & 0.321\% \\
\cline{1-5}
\bottomrule
\end{tabular}
        }
        \captionof{table}{Particle type distribution. Approximately 97.8\% of the examples belong to the 6 most populous particle types.}
        \label{tab:part_type_distr}
    \end{minipage}
    \hfill
    \begin{minipage}[b]{.45\textwidth}
      \centering
      \includegraphics[width=\linewidth]{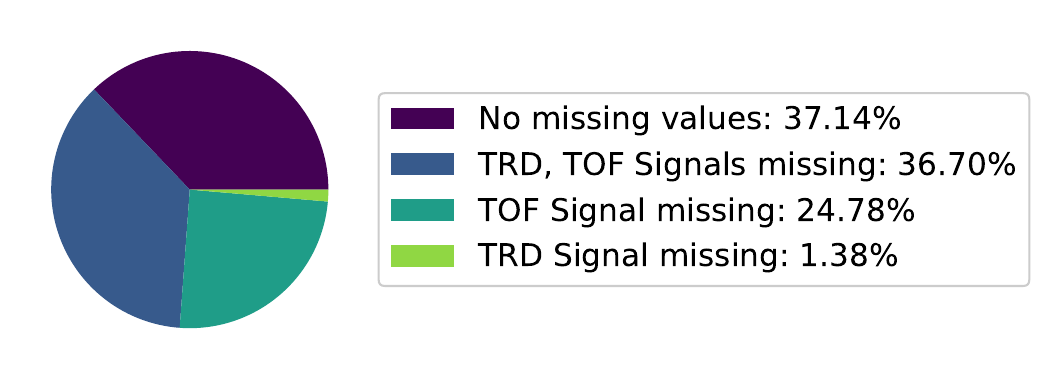}
      \captionof{figure}{Missing data distribution. Over 62.8\% of the examples are missing at least one value.}
      \label{fig:miss_val_distr}
    \end{minipage}
\end{figure}

The dataset consists of 2 751 934 examples with track transverse momentum \mbox{$\pt \geq 0.1$ \GeVc}. 95\% of examples fall into the range $[0.12, 1.76]$ \GeVc.
% Each example contains 19 features, representing signals from TPC, TRD, and TOF detectors as well as predetermined properties such as spatial coordinates of a track reconstruction starting point ($x$, $y$, $z$, angle $\alpha$), track momentum, type, charge, and the distances of closest approach (DCA) of the track trajectory to the collision's primary vertex measured in the $xy$ plane ($d_{xy}$) and the $z$ direction ($d_{z}$).
Each example contains 19 features:
\begin{itemize}
    \item detector signals: 1 TPC feature, 2 features per TRD and TOF
    \item number of shared TPC clusters
    \item spatial coordinates of a track reconstruction starting point ($x$, $y$, $z$ in the local coordinate system, and the rotation angle $\alpha$ between local and global coordinate systems),
    \item track momentum $p$ and its \pt, $p_x$, $p_y$, $p_z$ components,
    \item track charge and type (propagated/non-propagated Run 3 track, Run 2 track)
    \item the distances of closest approach (DCA) of the track trajectory to the collision primary vertex measured in the $xy$ plane ($d_{xy}$) and the $z$ direction ($d_{z}$)
\end{itemize}
The examples are labeled with ten different particle types. The particle species distribution is shown in Table \ref{tab:part_type_distr}.

There are four combinations of missing values, as some examples' measurements from the TOF and TRD detectors are missing. Figure \ref{fig:miss_val_distr} shows the missing value distribution.

\subsection{Models}
For each compared method, we train identical binary classification models for 6 of the most populous particle types: pions, protons, kaons, and their respective antiparticles.

All trained models are identical for imputation methods and the neural network ensemble. They have three hidden layers of sizes 64, 32 and 16, and a single output.
Between the layers, we use Rectified Linear Unit (ReLU) activation function $f(x) = \max({0, x})$. Dropout regularization with a rate of 0.1 is applied after each activation layer.
The input dimension depends on the method used. For imputation techniques, all models process inputs of size 19, as all missing features are imputed.
For the neural net ensemble, the four networks have input sizes 19, 17, 17, and 15, depending on the combination of missing values from the TRD and TOF detectors (each detector has two features associated with it).

Our proposed architecture consists of the embedding neural network, the Transformer's encoder, the self-attention neural network, and the classifier network. 
ReLU activation is also used between neural network layers, and dropout regularization with a rate of 0.1 is applied to the output of the embedding layer and the output of each sub-layer in the encoder.
The parameters of all the layers are shown in Table~\ref{our_params}.

\begin{table}
\centering
\caption{Training hyperparameters.}
\label{our_params}
\resizebox{\linewidth}{!}{%
\begin{tabular}{ccccccccccccccc} 
\hline
\multicolumn{3}{c}{\multirow{3}{*}{Embedding}} & \multicolumn{6}{c}{Transformer Encoder} & \multicolumn{3}{c}{\multirow{3}{*}{Self-Attention}} & \multicolumn{3}{c}{\multirow{3}{*}{Classifier}} \\ 
% \vspace{-1pt}
\cmidrule{4-9}
\multicolumn{3}{c}{} & \multicolumn{5}{c}{Encoder Layer} & \multirow{3}{*}{Num Layers} & \multicolumn{3}{c}{} & \multicolumn{3}{c}{} \\ 
\cmidrule{4-8}
\multicolumn{3}{c}{} & \multicolumn{2}{c}{Multi-Head Attention} & \multicolumn{3}{c}{Neural Network} &  & \multicolumn{3}{c}{} & \multicolumn{3}{c}{} \\ 
\cline{1-8}\cline{10-15}
Input & Hidden & Output & Dimension & Heads & Input & Hidden & Output &  & Input & Hidden & Output & Input & Hidden & Output \\ 
\hline
\multicolumn{1}{r}{20} & \multicolumn{1}{r}{128} & \multicolumn{1}{r}{32} & \multicolumn{1}{r}{32} & \multicolumn{1}{r}{2} & \multicolumn{1}{r}{32} & \multicolumn{1}{r}{128} & \multicolumn{1}{r}{32} & \multicolumn{1}{r}{2} & \multicolumn{1}{r}{32} & \multicolumn{1}{r}{64} & \multicolumn{1}{r}{32} & \multicolumn{1}{r}{32} & \multicolumn{1}{r}{64} & \multicolumn{1}{r}{1} \\ 
\hline
\multicolumn{1}{l}{} & \multicolumn{1}{l}{} & \multicolumn{1}{l}{} & \multicolumn{1}{l}{} & \multicolumn{1}{l}{} & \multicolumn{1}{l}{} & \multicolumn{1}{l}{} & \multicolumn{1}{l}{} & \multicolumn{1}{l}{} & \multicolumn{1}{l}{} & \multicolumn{1}{l}{} & \multicolumn{1}{l}{} & \multicolumn{1}{l}{} & \multicolumn{1}{l}{} & \multicolumn{1}{l}{}
\end{tabular}
}
\end{table}

Model parameters are selected using a hyperparameter sweep for a fair comparison of different network architectures. Given a set of parameter, various combinations of their values are used to train different models, from which the model achieving the best results on the validation dataset is chosen. In this way, each of the compared methods may achieve the best possible results. The hyperparameter sweep is a computationally expensive procedure, hence we perform it with a model trained to detect kaons -- the most challenging particles to identify in our dataset. This might bring a small bias of our results towards kaons, so before the full integration of our solution with the ALICE system, we will perform independent sweeps for all particles. %The parameter combinations are evaluated on the kaon classification task using the proposed architecture and the standard feed-forward neural network. 

\subsection{Results}
Each model is trained using all complete and incomplete examples and tested in two cases: (1) with all available test examples and (2) with the complete ones only. 
For case deletion, no results are available for testing on the missing data dataset, since it is not possible to directly apply the appropriate models on data with missing values. We evaluate our models on test-set with standard metrics -- precision (purity) and recall (efficiency). Since precision and recall are antagonistic, we also include the $F_1$ metric, which is a combination of the first two according to the formula:
\begin{equation}
    F_1=\frac{2 \cdot precision~\cdot~recall}{precision~+~recall}.
\end{equation}
Numerical results for the pion, proton, and kaon particles are shown in table \ref{tab:results_table} and in table~\ref{tab:results_antiparticles_table} for antiparticles. We highlight the best-performing method for each particle in bold.

\begin{table}[H]
    \def\arraystretch{1.3}
    \caption{Classification result for the three most common particle species.}
    \label{tab:results_table}
   
    \begin{subtable}{0.32\textwidth}
        \caption{Pion identification on complete data only.}
        \resizebox{\linewidth}{!}{
            \begin{tabular}{llll}
\toprule
Model & Precision & Recall & $F_1$ \\
% model &  &  &  \\
\midrule
Delete & 99.08 $\pm$ 0.07 & 99.67 $\pm$ 0.04 & 99.37 $\pm$ 0.01 \\
\cline{1-4}
Ensemble & \textbf{99.11 $\pm$ 0.04} & 99.64 $\pm$ 0.06 & \textbf{99.38 $\pm$ 0.01} \\
\cline{1-4}
Mean & 98.85 $\pm$ 0.09 & \textbf{99.69 $\pm$ 0.04} & 99.27 $\pm$ 0.04 \\
\cline{1-4}
Proposed & 99.08 $\pm$ 0.02 & 99.64 $\pm$ 0.03 & 99.36 $\pm$ 0.01 \\
\cline{1-4}
Regression & 99.02 $\pm$ 0.02 & 99.49 $\pm$ 0.14 & 99.25 $\pm$ 0.07 \\
\cline{1-4}
\bottomrule
\end{tabular}

        }
    \end{subtable}
    \hfill
    \begin{subtable}{0.32\textwidth}
        \caption{Proton identification on complete data only.}
        \resizebox{\linewidth}{!}{
            \begin{tabular}{llll}
\toprule
Model & Precision & Recall & $F_1$ \\
\midrule
Delete & 99.23 $\pm$ 0.32 & 99.63 $\pm$ 0.05 & 99.43 $\pm$ 0.16 \\
\cline{1-4}
Ensemble & 99.16 $\pm$ 0.24 & \textbf{99.76 $\pm$ 0.07} & 99.46 $\pm$ 0.13 \\
\cline{1-4}
Mean & 99.22 $\pm$ 0.19 & 99.72 $\pm$ 0.04 & 99.47 $\pm$ 0.08 \\
\cline{1-4}
Proposed & \textbf{99.28 $\pm$ 0.10} & 99.68 $\pm$ 0.09 & \textbf{99.48 $\pm$ 0.02} \\
\cline{1-4}
Regression & 99.10 $\pm$ 0.09 & 99.65 $\pm$ 0.09 & 99.37 $\pm$ 0.07 \\
\cline{1-4}
\bottomrule
\end{tabular}

        }
    \end{subtable}
    \hfill
    \begin{subtable}{0.32\textwidth}
        \caption{Kaon identification on complete data only.}
        \resizebox{\linewidth}{!}{
            \begin{tabular}{llll}
\toprule
Model & Precision & Recall & $F_1$ \\
\midrule
Delete & \textbf{96.93 $\pm$ 0.37} & 96.98 $\pm$ 0.26 & 96.95 $\pm$ 0.06 \\
\cline{1-4}
Ensemble & 96.65 $\pm$ 0.38 & 97.82 $\pm$ 0.31 & \textbf{97.23 $\pm$ 0.10} \\
\cline{1-4}
Mean & 96.83 $\pm$ 0.17 & 95.33 $\pm$ 0.67 & 96.08 $\pm$ 0.36 \\
\cline{1-4}
Proposed & 96.03 $\pm$ 0.98 & \textbf{98.06 $\pm$ 0.72} & 97.04 $\pm$ 0.17 \\
\cline{1-4}
Regression & 94.27 $\pm$ 0.98 & 97.01 $\pm$ 0.51 & 95.62 $\pm$ 0.39 \\
\cline{1-4}
\bottomrule
\end{tabular}

        }
    \end{subtable}
    \vspace{5mm}

    \begin{subtable}{0.32\textwidth}
        \caption{Pion identification on data including incomplete examples.}
        \resizebox{\linewidth}{!}{
            \begin{tabular}{llll}
\toprule
Model & Precision & Recall & $F_1$ \\
\midrule
Standard & \textbf{99.99 $\pm$ 0.01} & 78.37 $\pm$ 0.01 & 87.87 $\pm$ 0.87\\
\midrule
\midrule
Ensemble & 97.47 $\pm$ 0.25 & 99.46 $\pm$ 0.21 & 98.45 $\pm$ 0.04 \\
\cline{1-4}
Mean & 97.31 $\pm$ 0.07 & 99.52 $\pm$ 0.07 & 98.40 $\pm$ 0.01 \\
\cline{1-4}
Proposed & 97.49 $\pm$ 0.06 & \textbf{99.54 $\pm$ 0.05} & \textbf{98.50 $\pm$ 0.02} \\
\cline{1-4}
Regression & 97.33 $\pm$ 0.06 & 99.49 $\pm$ 0.07 & 98.40 $\pm$ 0.04 \\
\cline{1-4}
\bottomrule
\end{tabular}

        }
    \end{subtable}
    \hfill
    \begin{subtable}{0.32\textwidth}
        \caption{Proton identification on data including incomplete examples.}
        \resizebox{\linewidth}{!}{
            \begin{tabular}{llll}
\toprule
Model & Precision & Recall & $F_1$ \\
\midrule
Standard &  \textbf{99.40 $\pm$ 0.01} & 59.72 $\pm$ 0.03 & 74.61 $\pm$ 1.88\\
\midrule
\midrule
Ensemble & 97.16 $\pm$ 0.46 & 93.74 $\pm$ 0.30 & 95.42 $\pm$ 0.12 \\
\cline{1-4}
Mean & 97.85 $\pm$ 0.41 & 93.34 $\pm$ 0.32 & 95.54 $\pm$ 0.06 \\
\cline{1-4}
Proposed & 97.80 $\pm$ 0.44 & \textbf{93.86 $\pm$ 0.27} & \textbf{95.79 $\pm$ 0.07} \\
\cline{1-4}
Regression & 97.38 $\pm$ 0.40 & 93.67 $\pm$ 0.38 & 95.49 $\pm$ 0.15 \\
\cline{1-4}
\bottomrule
\end{tabular}

        }
    \end{subtable}
    \hfill
    \begin{subtable}{0.32\textwidth}
         \caption{Kaon identification on data including incomplete examples.}
        \resizebox{\linewidth}{!}{
            \begin{tabular}{llll}
\toprule
Model & Precision & Recall & $F_1$ \\
\midrule
Standard & \textbf{92.87 $\pm$ 0.01} & 60.37 $\pm$ 0.05 & 73.17 $\pm$ 1.57 \\
\midrule
\midrule
Ensemble & 91.18 $\pm$ 02.00 & 82.72 $\pm$ 01.42 & 86.74 $\pm$ 0.16 \\
\cline{1-4}
Mean & 90.83 $\pm$ 01.71 & 82.32 $\pm$ 0.96 & 86.36 $\pm$ 0.34 \\
\cline{1-4}
Proposed & 91.55 $\pm$ 0.71 & \textbf{83.68 $\pm$ 0.82} & \textbf{87.44 $\pm$ 0.14} \\
\cline{1-4}
Regression & 91.17 $\pm$ 01.00 & 81.78 $\pm$ 0.21 & 86.22 $\pm$ 0.46 \\
\cline{1-4}
\bottomrule
\end{tabular}

        }
    \end{subtable}
\end{table}

\begin{table}[H]
    \def\arraystretch{1.3}
    \caption{Classification result for the three most common antiparticle species.}
     \label{tab:results_antiparticles_table}
    \begin{subtable}{0.32\textwidth}
        \caption{Antipion identification on complete data only.}
        \resizebox{\linewidth}{!}{
            \begin{tabular}{llll}
\toprule
Model & Precision & Recall & $F_1$ \\
\midrule
Delete & \textbf{99.08 $\pm$ 0.04} & 99.67 $\pm$ 0.02 & \textbf{99.37 $\pm$ 0.01} \\
\cline{1-4}
Ensemble & 98.93 $\pm$ 0.51 & \textbf{99.76 $\pm$ 0.16} & 99.34 $\pm$ 0.18 \\
\cline{1-4}
Mean & 98.86 $\pm$ 0.11 & 99.69 $\pm$ 0.03 & 99.27 $\pm$ 0.04 \\
\cline{1-4}
Proposed & \textbf{99.08 $\pm$ 0.04} & 99.67 $\pm$ 0.02 & \textbf{99.37 $\pm$ 0.03} \\
\cline{1-4}
Regression & 99.04 $\pm$ 0.03 & 99.51 $\pm$ 0.05 & 99.28 $\pm$ 0.02 \\
\cline{1-4}
\bottomrule
\end{tabular}

        }
    \end{subtable}
    \hfill
    \begin{subtable}{0.32\textwidth}
        \caption{Antiproton identification on complete data only.}
        \resizebox{\linewidth}{!}{
            \begin{tabular}{llll}
\toprule
Model & Precision & Recall & $F_1$ \\
\midrule
Delete & 98.75 $\pm$ 0.37 & 99.52 $\pm$ 0.17 & 99.13 $\pm$ 0.26 \\
\cline{1-4}
Ensemble & 99.12 $\pm$ 0.08 & 99.53 $\pm$ 0.16 & 99.33 $\pm$ 0.10 \\
\cline{1-4}
Mean & 98.79 $\pm$ 0.58 & 99.62 $\pm$ 0.18 & 99.20 $\pm$ 0.27 \\
\cline{1-4}
Proposed & \textbf{99.25 $\pm$ 0.06} & 99.63 $\pm$ 0.20 & \textbf{99.44 $\pm$ 0.08} \\
\cline{1-4}
Regression & 98.57 $\pm$ 0.33 & \textbf{99.64 $\pm$ 0.25} & 99.10 $\pm$ 0.13 \\
\cline{1-4}
\bottomrule
\end{tabular}

        }
    \end{subtable}
    \hfill
    \begin{subtable}{0.32\textwidth}
        \caption{Antikaon identification on complete data only.}
        \resizebox{\linewidth}{!}{
            \begin{tabular}{llll}
\toprule
Model & Precision & Recall & $F_1$ \\
\midrule
Delete & 95.82 $\pm$ 0.69 & 96.84 $\pm$ 0.66 & 96.33 $\pm$ 0.11 \\
\cline{1-4}
Ensemble & \textbf{96.14 $\pm$ 0.32} & 97.60 $\pm$ 0.24 & 96.87 $\pm$ 0.09 \\
\cline{1-4}
Mean & \textbf{96.14 $\pm$ 0.45} & 94.77 $\pm$ 0.99 & 95.45 $\pm$ 0.33 \\
\cline{1-4}
Proposed & 96.00 $\pm$ 0.11 & \textbf{97.85 $\pm$ 0.12} & \textbf{96.91 $\pm$ 0.11} \\
\cline{1-4}
Regression & 93.85 $\pm$ 01.11 & 96.41 $\pm$ 0.18 & 95.11 $\pm$ 0.58 \\
\cline{1-4}
\bottomrule
\end{tabular}

        }
    \end{subtable}
    \vspace{5mm}

    \begin{subtable}{0.32\textwidth}
        \caption{Antipion identification on data including incomplete examples.}
        \resizebox{\linewidth}{!}{
            \begin{tabular}{llll}
\toprule
Model & Precision & Recall & $F_1$ \\
\midrule
Standard & \textbf{99.99 $\pm$ 0.01} & 78.03 $\pm$ 0.01 & 87.66 $\pm$ 0.87 \\
\midrule
\midrule
Ensemble & 97.01 $\pm$ 0.87 & \textbf{99.56 $\pm$ 0.11} & 98.27 $\pm$ 0.42 \\
\cline{1-4}
Mean & 97.23 $\pm$ 0.03 & 99.47 $\pm$ 0.03 & 98.34 $\pm$ 0.01 \\
\cline{1-4}
Proposed & 97.38 $\pm$ 0.04 & 99.51 $\pm$ 0.02 & \textbf{98.44 $\pm$ 0.02} \\
\cline{1-4}
Regression & 97.22 $\pm$ 0.15 & 99.52 $\pm$ 0.12 & 98.36 $\pm$ 0.03 \\
\cline{1-4}
\bottomrule
\end{tabular}

        }
    \end{subtable}
    \hfill
    \begin{subtable}{0.32\textwidth}
        \caption{Antiproton identification on data including incomplete examples.}
        \resizebox{\linewidth}{!}{
            \begin{tabular}{llll}
\toprule
Model & Precision & Recall & $F_1$ \\
\midrule
Standard & \textbf{99.24 $\pm$ 0.01}& 53.02 $\pm$ 0.03 & 69.12 $\pm$ 1.93 \\
\midrule
\midrule
Ensemble & 96.90 $\pm$ 0.24 & \textbf{92.41 $\pm$ 0.21} & 94.60 $\pm$ 0.10 \\
\cline{1-4}
Mean & 97.24 $\pm$ 0.39 & 92.39 $\pm$ 0.15 & 94.75 $\pm$ 0.20 \\
\cline{1-4}
Proposed & 97.51 $\pm$ 0.55 & 92.40 $\pm$ 0.74 & \textbf{94.89 $\pm$ 0.14} \\
\cline{1-4}
Regression & 96.89 $\pm$ 0.50 & 92.36 $\pm$ 0.22 & 94.57 $\pm$ 0.13 \\
\cline{1-4}
\bottomrule
\end{tabular}

        }
    \end{subtable}
    \hfill
    \begin{subtable}{0.32\textwidth}
        \caption{Antikaon identification on data including incomplete examples.}
        \resizebox{\linewidth}{!}{
            \begin{tabular}{llll}
\toprule
Model & Precision & Recall & $F_1$ \\
\midrule
Standard &\textbf{92.22 $\pm$ 0.01}& 55.68 $\pm$ 0.04 & 69.44 $\pm$ 1.60 \\
\midrule
\midrule
Ensemble & 89.16 $\pm$ 01.51 & 81.06 $\pm$ 1.74 & 84.91 $\pm$ 0.48 \\
\cline{1-4}
Mean & 89.75 $\pm$ 0.80 & 80.14 $\pm$ 1.18 & 84.67 $\pm$ 0.38 \\
\cline{1-4}
Proposed & 90.86 $\pm$ 0.70 & \textbf{81.64 $\pm$ 0.63} & \textbf{86.00 $\pm$ 0.13} \\
\cline{1-4}
Regression & 91.63 $\pm$ 0.58 & 79.29 $\pm$ 0.59 & 85.01 $\pm$ 0.13 \\
\cline{1-4}
\bottomrule
\end{tabular}

        }
    \end{subtable}
\end{table}

For the case of incomplete examples, we compare our machine learning solutions with the standard technique described in Section~\ref{sec:pid}. For our comparison we used the following selections: $|n_{\rm \sigma,TPC}|<3$ for particles with transverse momenta below 0.5 \GeVc~and $\sqrt{n_{\rm \sigma,TPC}^{2}+n_{\rm \sigma,TOF}^{2}}<3$ for particles with $p_{\rm T} \geq 0.5$ \GeVc~(in this case TOF signal was required). Tables~\ref{tab:results_table} and~\ref{tab:results_antiparticles_table} show that machine learning approaches, in general, outperform standard $n_\sigma$-based techniques, providing significantly higher recall for similar or higher precision. Moreover, the proposed architecture is comparable with other tested techniques.
We can also observe that training with additional examples with missing data still results in the good quality of the PID on complete examples, as measured by $F_1$ scores.
On the other hand, synthetic imputation of mean or predicted missing values can disturb the learned function and might result in lower performance.
The cases of pion and kaon identification on complete cases are the only two where the ensemble achieves slightly better $F_1$ results than the proposed architecture. 

\subsection{Detailed analysis of the results}
\begin{figure}[h!]
\begin{minipage}[t]{.37\textwidth}
      \centering
      \includegraphics[width=\linewidth]{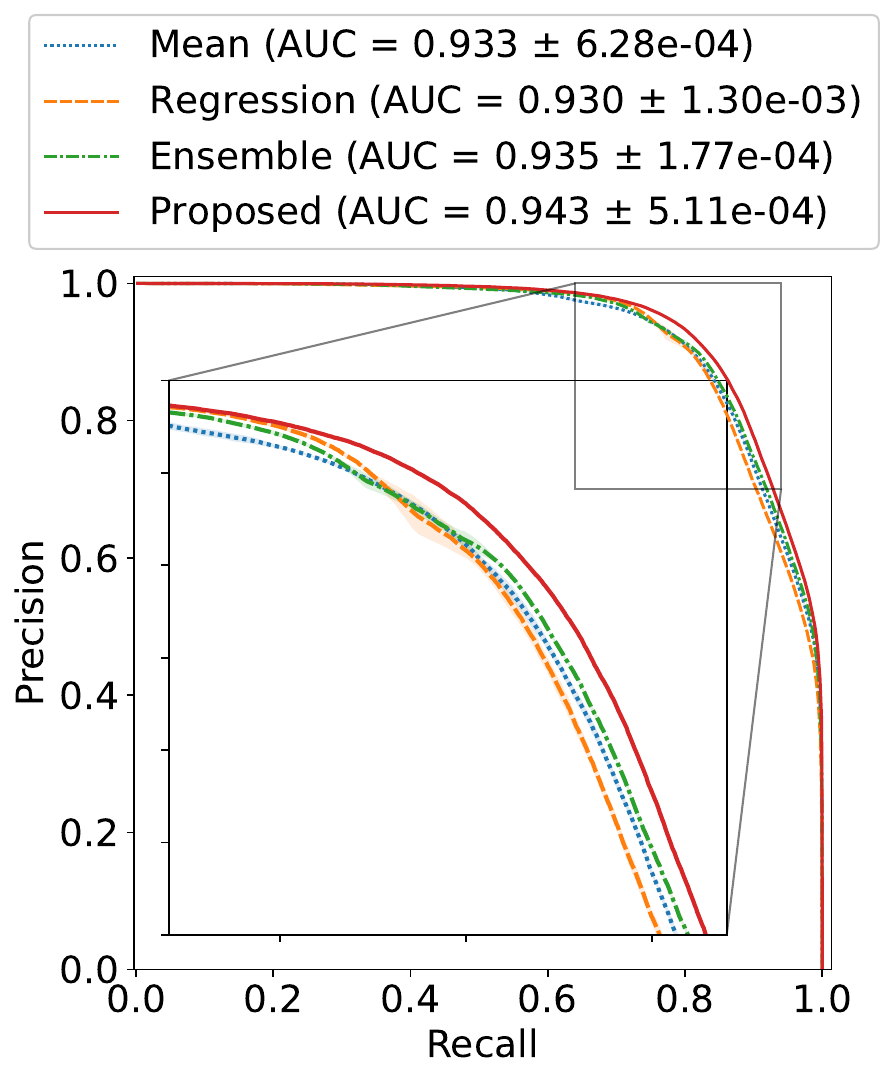}
      \captionof{figure}{Precision recall curve for different ML based approaches with missing data}
      \label{fig:prc_kaon_all}
    \end{minipage}
    \hfill
        \begin{minipage}[t]{.58\textwidth}
      \centering
      \includegraphics[width=\linewidth]{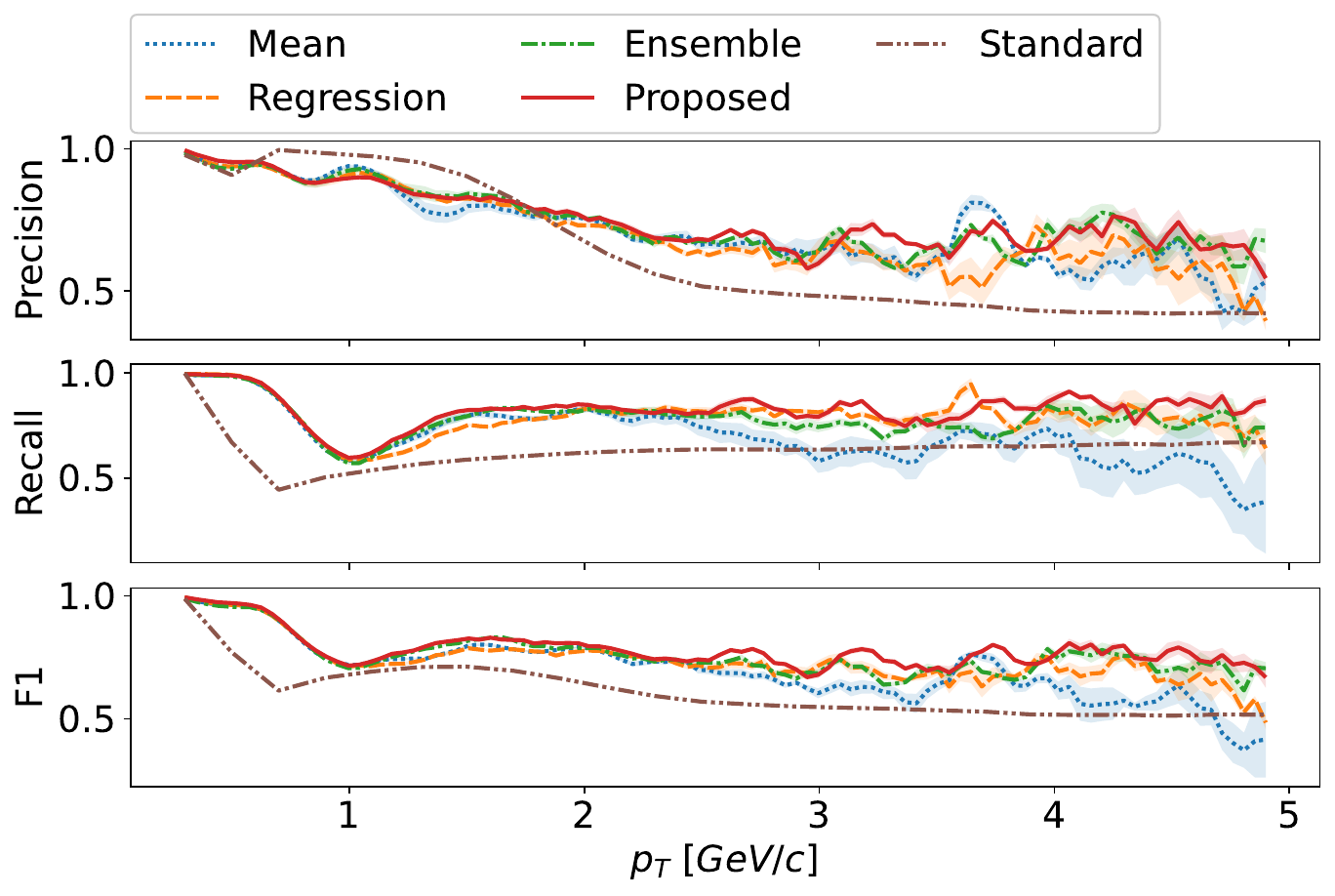}
      \captionof{figure}{Performance of different PID methods in kaon selection task with missing data as a function of particle momentum.}
      \label{fig:pt_kaon_all}
    \end{minipage}
\end{figure}

To further highlight the benefits of our approach, we provide a detailed analysis of its performance when compared to the baseline approaches. In Figure~\ref{fig:prc_kaon_all}, we present precision-recall curves of different methods applied on all available test data (including incomplete examples) in the most challenging kaon detection task. This plot provides detailed comparison between methods without a need for threshold selection. Additionally, in Figure~\ref{fig:pt_kaon_all}, we analyze the differences in performance for different range of particle momentum $p_T$. We can observe that our approach yields significantly higher area under the curve performance, and has a lower degradation in performance given particles with high momentum values.

\begin{figure}[h!]
   \begin{minipage}[t]{.37\textwidth}
      \centering
      \includegraphics[width=\linewidth]{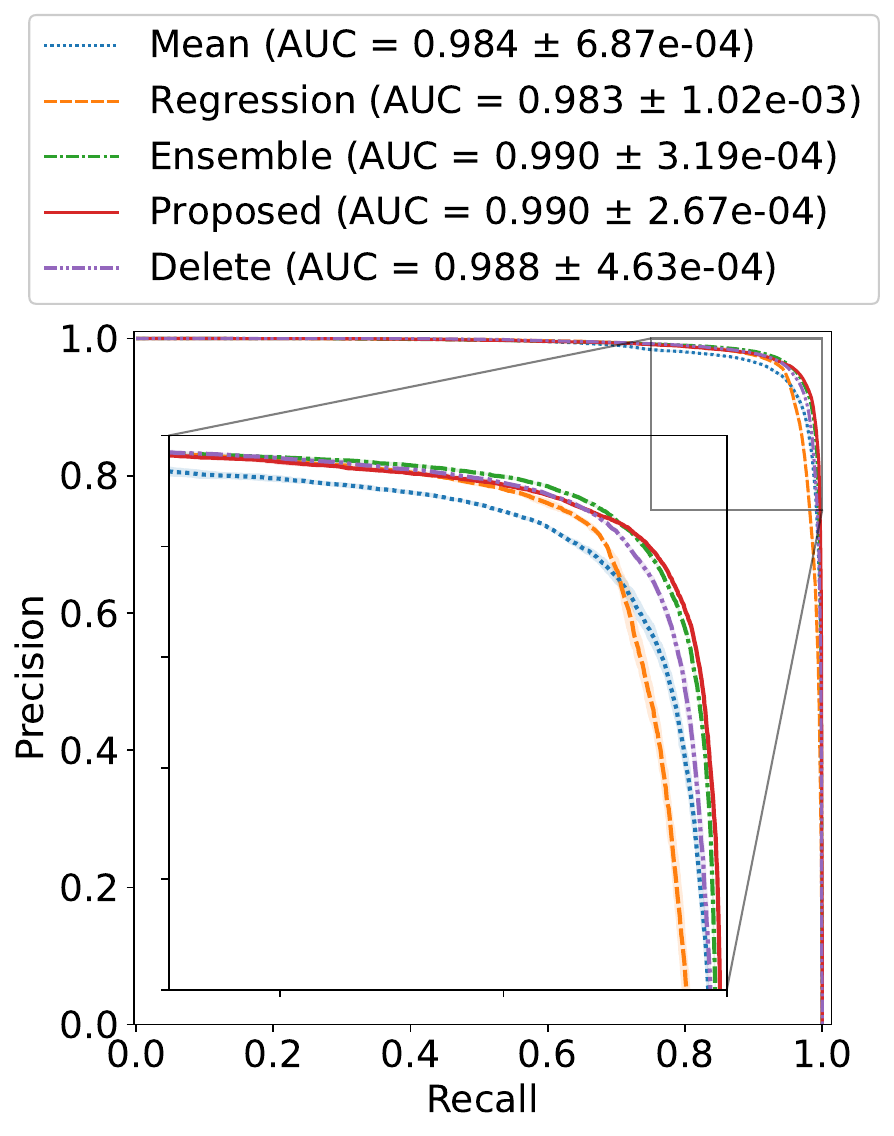}
      \captionof{figure}{Precision recall curve for different ML based approaches without missing data}
      \label{fig:prc_kaon_complete}
    \end{minipage} 
    \hfill
    \begin{minipage}[t]{.58\textwidth}
      \centering
      \includegraphics[width=\linewidth]{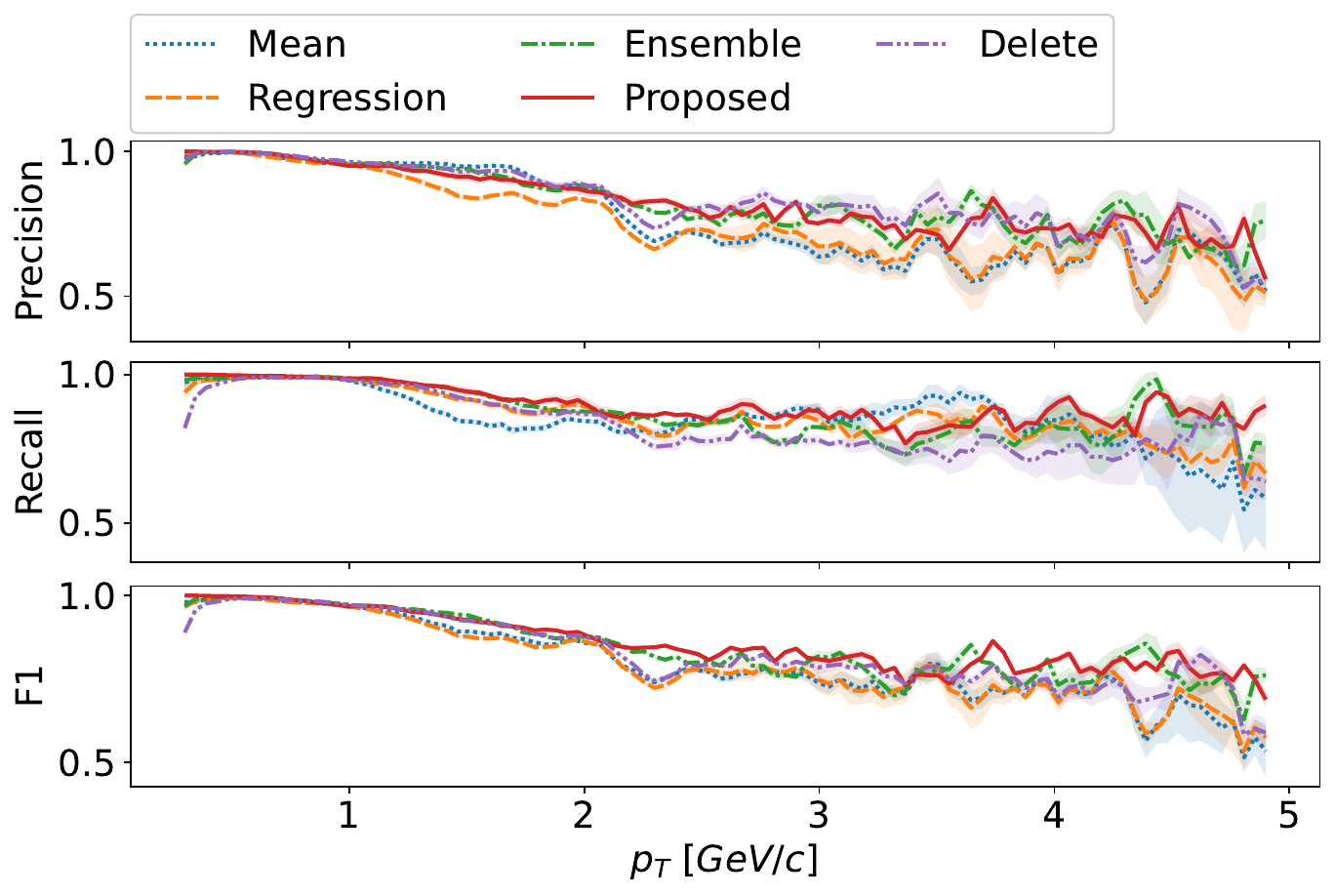}
      \captionof{figure}{Performance of different PID methods in kaon selection task without missing data as a function of particle momentum.}
      \label{fig:pt_kaon_complete}
    \end{minipage}
    
\end{figure}

In Figures~\ref{fig:prc_kaon_complete} and~\ref{fig:pt_kaon_complete}, we perform similar analysis for using only complete test data examples. Detailed evaluations for all other particles are included in the supplementary material.

\newpage
\subsection{Computational time comparison}

In Tables~\ref{tab:training_time} and~\ref{fig:inference_time}, we present the computational overhead of calculating predictions with our method, when compared to the other baselines. In particular for training time, we report average time needed (in milliseconds) to update model's weights from loading data, through forward and backward propagation up to the final update. For inference time, we report the average forward pass trough the network (in microseconds). Since our model requires calculation of the attention between features coming from different detectors, both it's training and inference times are around 3 times longer than for a standard methods. However, all methods can be easily parallelized (up to the GPU memory size), what can be seen in nearly constant times across different batch sizes. All of the calculations are performed on a small local machine with Nvidia GTX 1660Ti 6GB GPU, Intel core i7-9750H CPU, and 16 GB 2400 MHz (MT/s) of RAM.

\begin{figure}[h!]
    \begin{minipage}[b]{.45\textwidth}
        \centering
        \resizebox{\linewidth}{!}{
            \begin{tabular}{lrrrr}
            \toprule
            Batch size & 64 & 128 & 256 & 512  \\
            \midrule
            \midrule
            Mean & 3.10 & 3.54 & 3.69 & 3.64 \\
            \cline{1-5}
            Regression & 2.98 & 3.18 & 3.37 & 3.40 \\
            \cline{1-5}
            Ensemble & 3.67 & 3.62 & 3.74 & 3.99 \\
            \cline{1-5}
            Proposed & 11.09 & 10.21 & 10.82 & 12.65 \\
            \cline{1-5}
            Delete & 3.36 & 3.31 & 3.39 & 3.68 \\
            \cline{1-5}
            \bottomrule
            \end{tabular}
        }
        \captionof{table}{Average training time of our method compared to the baseline [$ms$]}
        \label{tab:training_time}
    \end{minipage}
    \hfill
    \begin{minipage}[b]{.49\textwidth}
      \centering
              \resizebox{\linewidth}{!}{
        \begin{tabular}{lrrrr}
        \toprule
        Batch size & 64 & 128 & 256 & 512 \\
        \midrule
        \midrule
        Mean & 413.1 & 364.0 & 355.7 & 365.3 \\
        \cline{1-5}
        Regression & 370.5 & 349.2 & 358.9 & 377.2 \\
        \cline{1-5}
        Ensemble & 415.1 & 411.6 & 436.8 & 420.7 \\
        \cline{1-5}
        Proposed & 1507.7 & 1561.9 & 1592.8 & 1530.8 \\
        \cline{1-5}
        Delete & 479.0 & 462.3 & 432.8 & 462.5 \\
        \cline{1-5}
        \bottomrule
        \end{tabular}
        }
      \captionof{table}{Average training time of our method compared to the baseline [$\mu s$]}
      \label{fig:inference_time}
    \end{minipage}
\end{figure}

% \newpage
% \newpage
\section{Conclusion}
This work considers the real-case scenario of classification from incomplete data in the particle identification task, where due to the nature of physical processes, not all of the information is always recorded by all of the detectors. To solve this problem, we propose a novel method based on the attention mechanism. We verify that our approach is able to learn from both complete and incomplete data improving the performance of models trained solely on the complete examples. Moreover, our method provides no worse performance than other techniques while avoiding their drawbacks such as insertion of artificial data (imputation methods) and potentially too complex architecture (neural network ensemble).

% Future research may compare the proposed method against other, more sophisticated methods for incomplete data classification.
% The method may also be expanded to simplify the processing of features that are always available, as this may drastically reduce the required computation in cases where only a few of the features in the data may be missing.

\section*{Acknowledgements}
We would like to thank the ALICE Collaboration for guidance and support during our research as well as for the access to all software and data.\\

This work was supported by the Polish National Science Centre under agreements no. no. 2021/43/D/ST2/02214 and UMO-2022/45/B/ST2/02029, by the Polish Ministry for Education and Science under agreements no. 2022/WK/01 and 5236/CERN/2022/0, as well as by the IDUB-POB-FWEiTE-2 project granted by Warsaw University of Technology under the program Excellence Initiative: Research University (ID-UB).

\clearpage

% % \bibliographystyle{unsrt}
\bibliographystyle{sn-mathphys}
\bibliography{refs}

% \bibliography{sn-bibliography}% common bib file

\end{document}

\section*{Supplementary material:
Machine-learning-based particle identification with missing data}
\section{Detailed evaluations for all particles}
In the following sections, we show results of the detailed comparison between our and benchmark approaches for all of the particles we consider in our PID task.
% \newpage
\subsection{Kaons}

\begin{figure}[h!]
  \vspace*{-1.1cm}
    \begin{minipage}[t]{.58\textwidth}
      \centering
      \includegraphics[width=\linewidth]{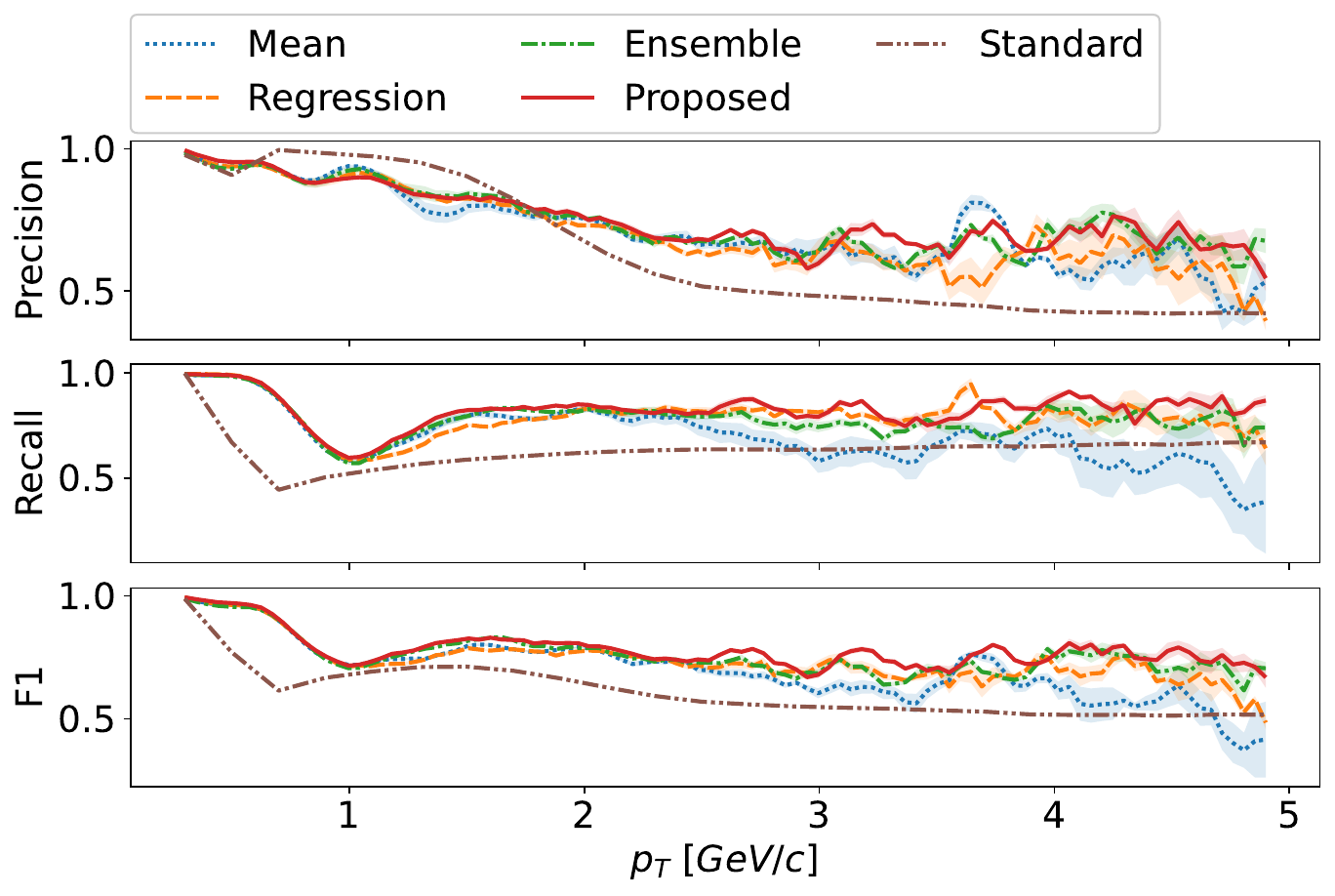}
      \captionof{figure}{Performance of different PID methods in kaon selection task with missing data as a function of particle momentum.}
      \label{fig:pt_kaon_all}
    \end{minipage}
    \hfill
    \begin{minipage}[t]{.37\textwidth}
      \centering
      \includegraphics[width=\linewidth]{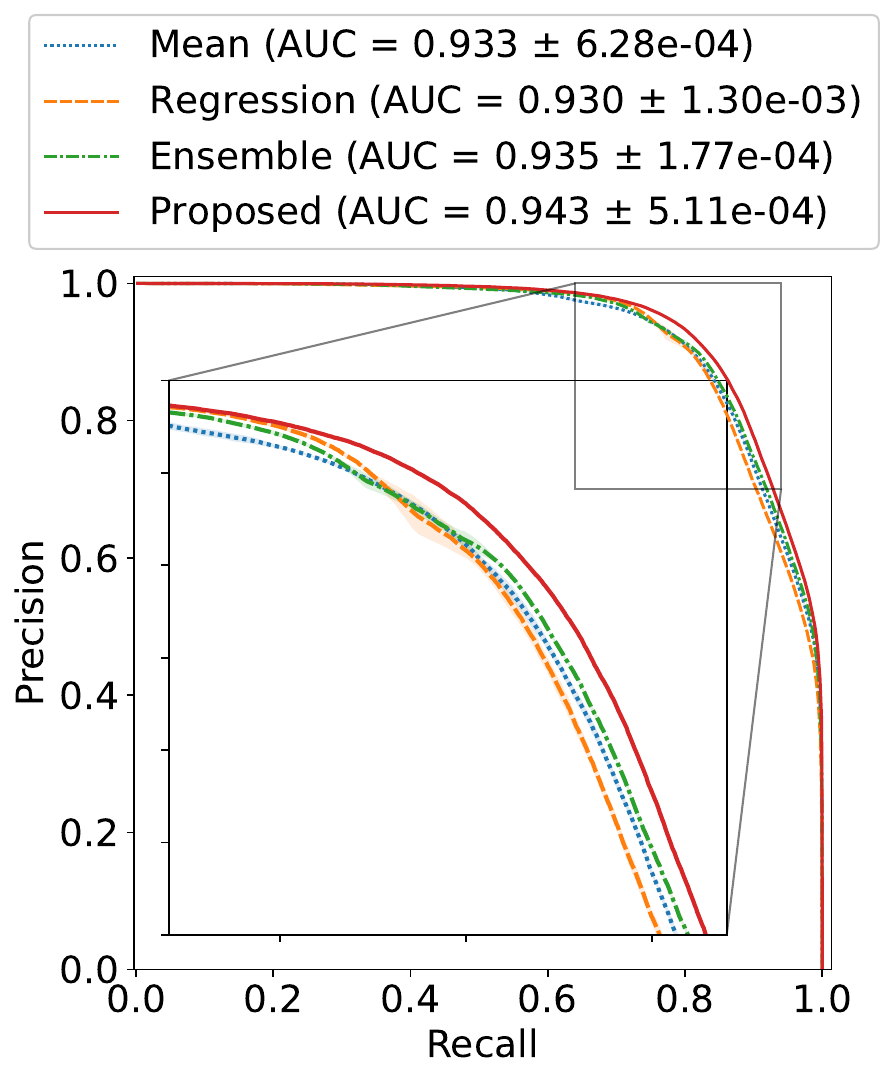}
      \captionof{figure}{Precision recall curve for different ML based approaches with missing data}
      \label{fig:prc_kaon_all}
    \end{minipage}
\end{figure}

\begin{figure}[h!]
    \vspace*{-1.1cm}
    \begin{minipage}[t]{.58\textwidth}
      \centering
      \includegraphics[width=\linewidth]{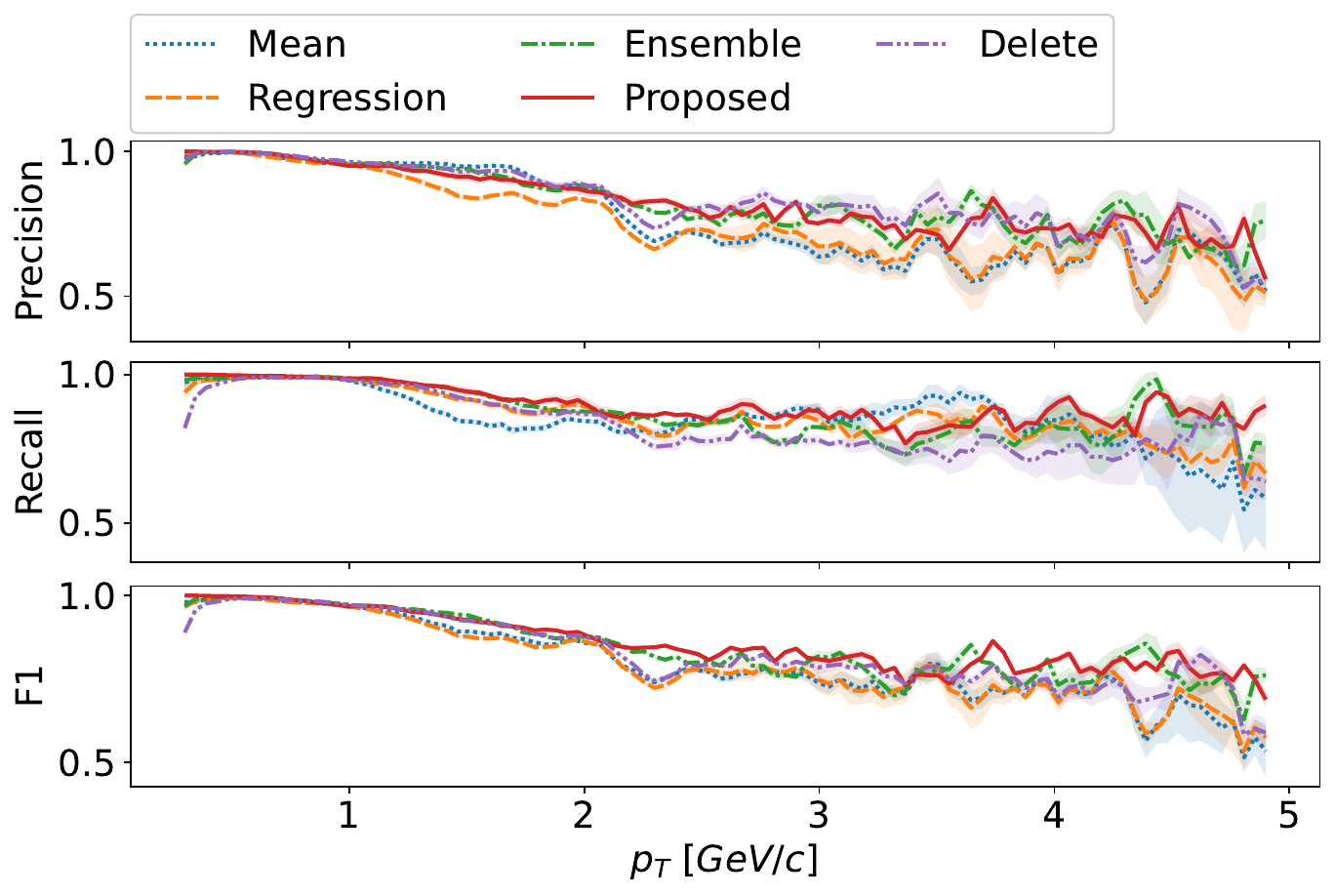}
      \captionof{figure}{Performance of different PID methods in kaon selection task without missing data as a function of particle momentum.}
      \label{fig:pt_kaon_complete}
    \end{minipage}
    \hfill
    \begin{minipage}[t]{.37\textwidth}
      \centering
      \includegraphics[width=\linewidth]{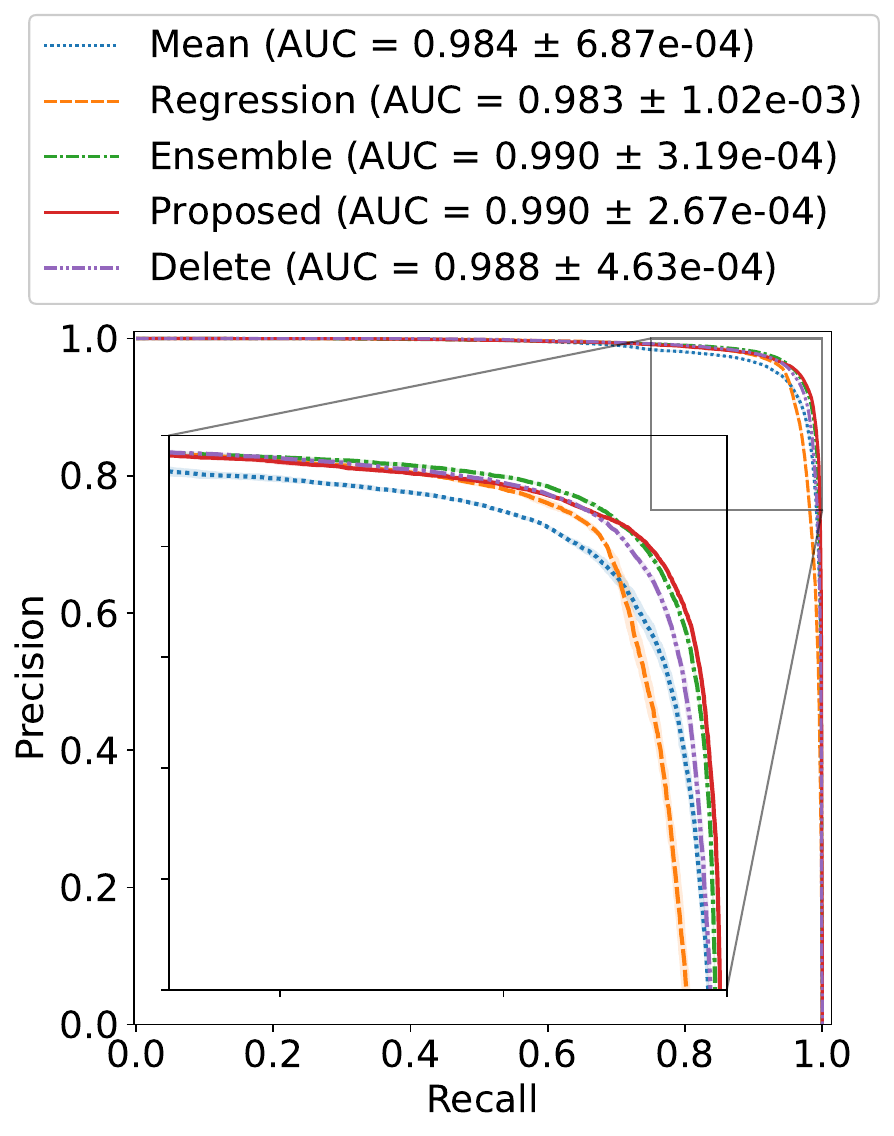}
      \captionof{figure}{Precision recall curve for different ML based approaches without missing data}
      \label{fig:prc_kaon_complete}
    \end{minipage}
\end{figure}

\newpage
\subsection{Protons}

\begin{figure}[h!]
    \begin{minipage}[t]{.58\textwidth}
      \centering
      \includegraphics[width=\linewidth]{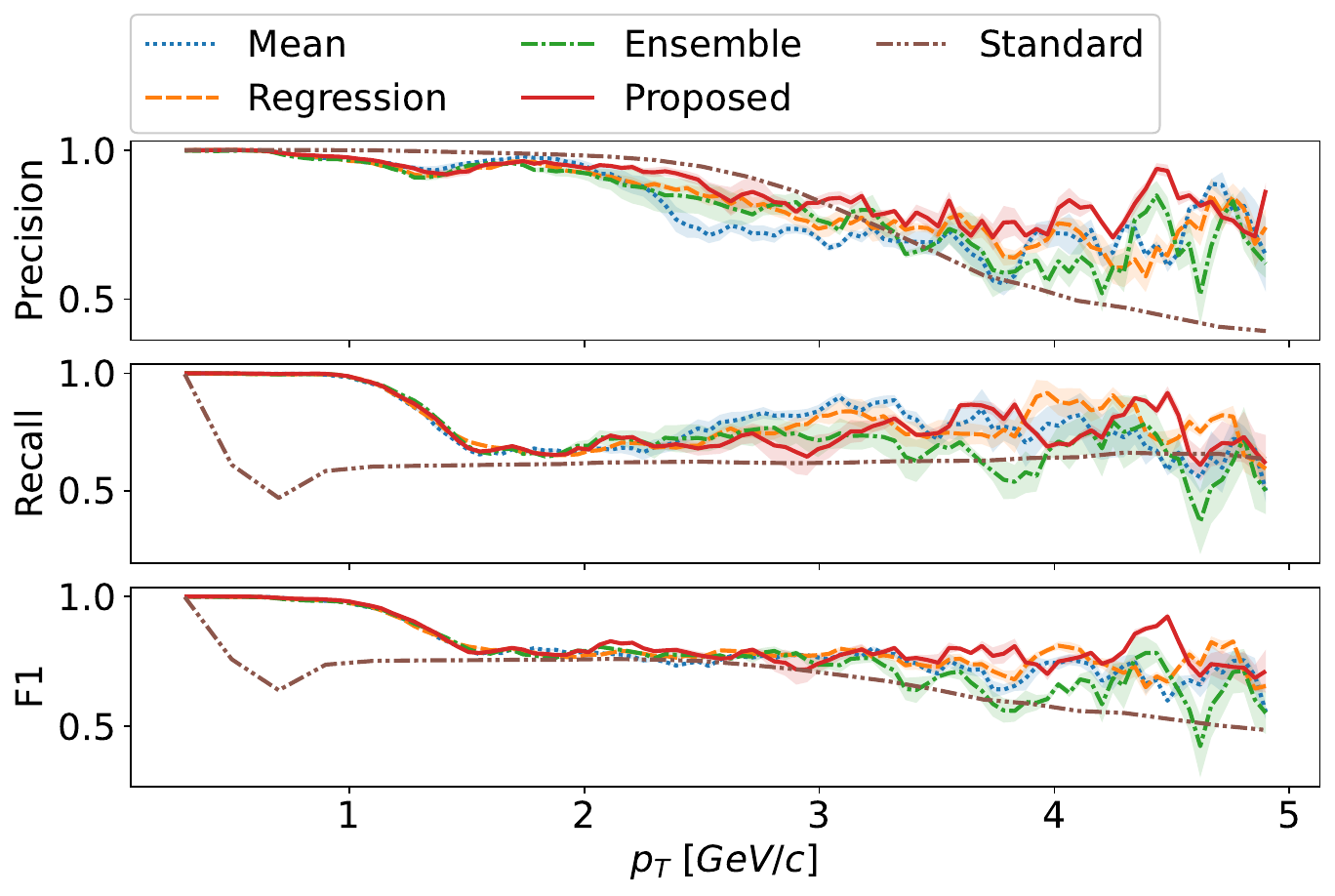}
      \captionof{figure}{Performance of different PID methods in proton selection task with missing data as a function of particle momentum.}
      \label{fig:pt_kaon_all}
    \end{minipage}
    \hfill
    \begin{minipage}[t]{.37\textwidth}
      \centering
      \includegraphics[width=\linewidth]{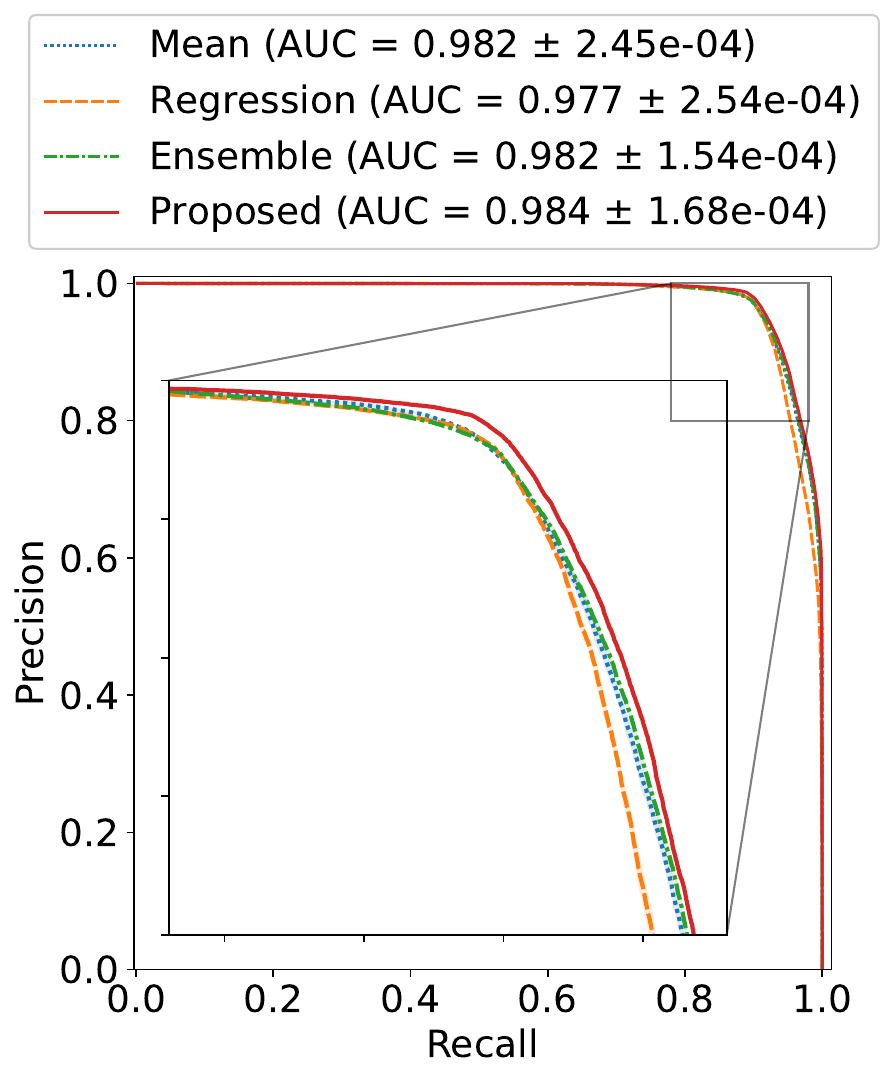}
      \captionof{figure}{Precision recall curve for different ML based approaches with missing data}
      \label{fig:prc_kaon_all}
    \end{minipage}
\end{figure}

\begin{figure}[h!]
    \begin{minipage}[t]{.58\textwidth}
      \centering
      \includegraphics[width=\linewidth]{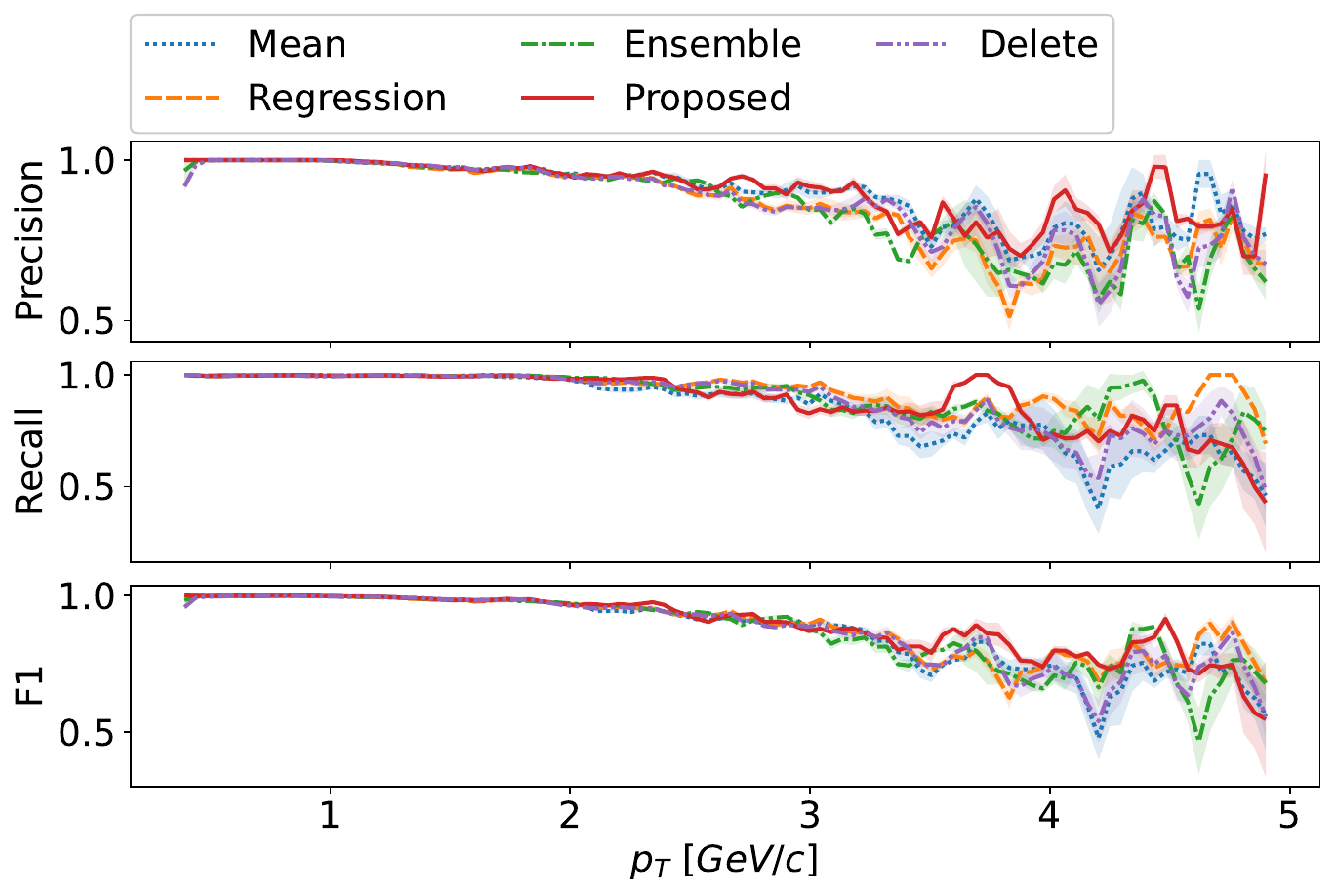}
      \captionof{figure}{Performance of different PID methods in proton selection task without missing data as a function of particle momentum.}
      \label{fig:pt_kaon_complete}
    \end{minipage}
    \hfill
    \begin{minipage}[t]{.37\textwidth}
      \centering
      \includegraphics[width=\linewidth]{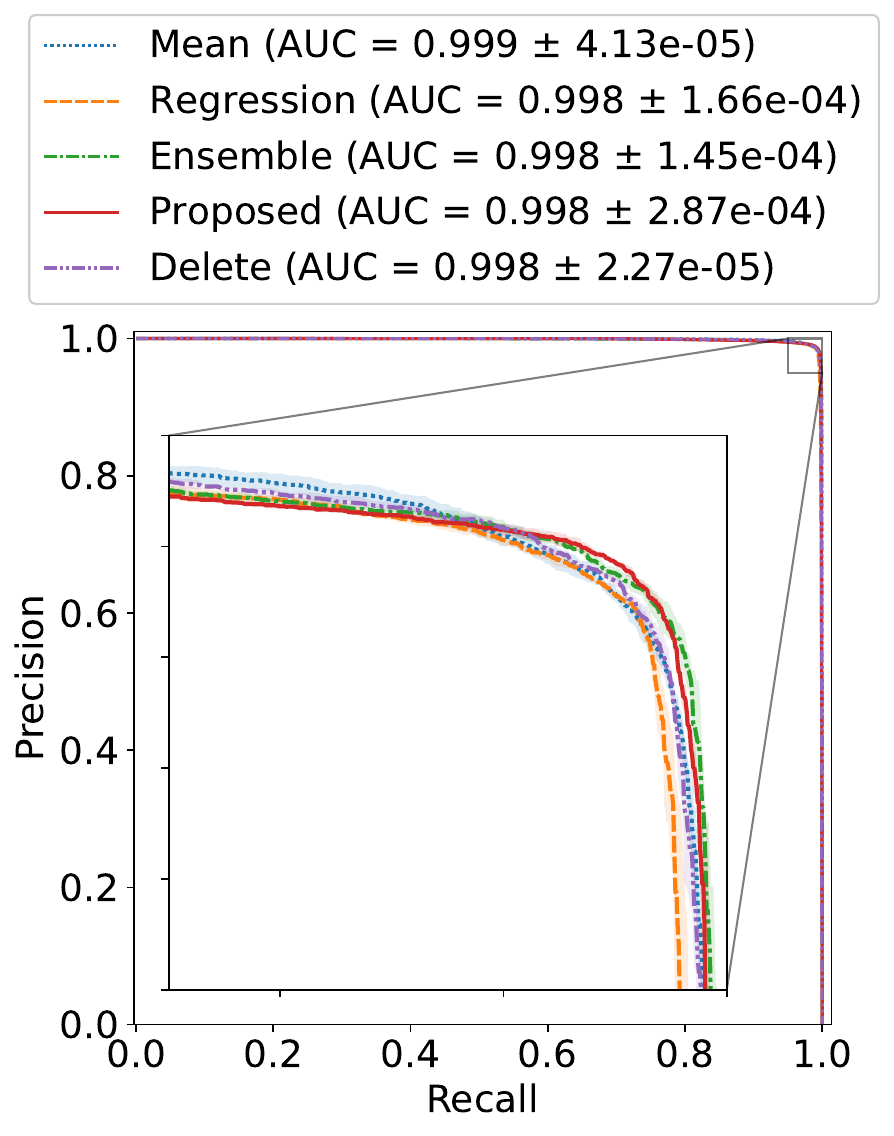}
      \captionof{figure}{Precision recall curve for different ML based approaches without missing data}
      \label{fig:prc_kaon_complete}
    \end{minipage}
\end{figure}

\newpage
\subsection{Pions}

\begin{figure}[h!]
    \begin{minipage}[t]{.58\textwidth}
      \centering
      \includegraphics[width=\linewidth]{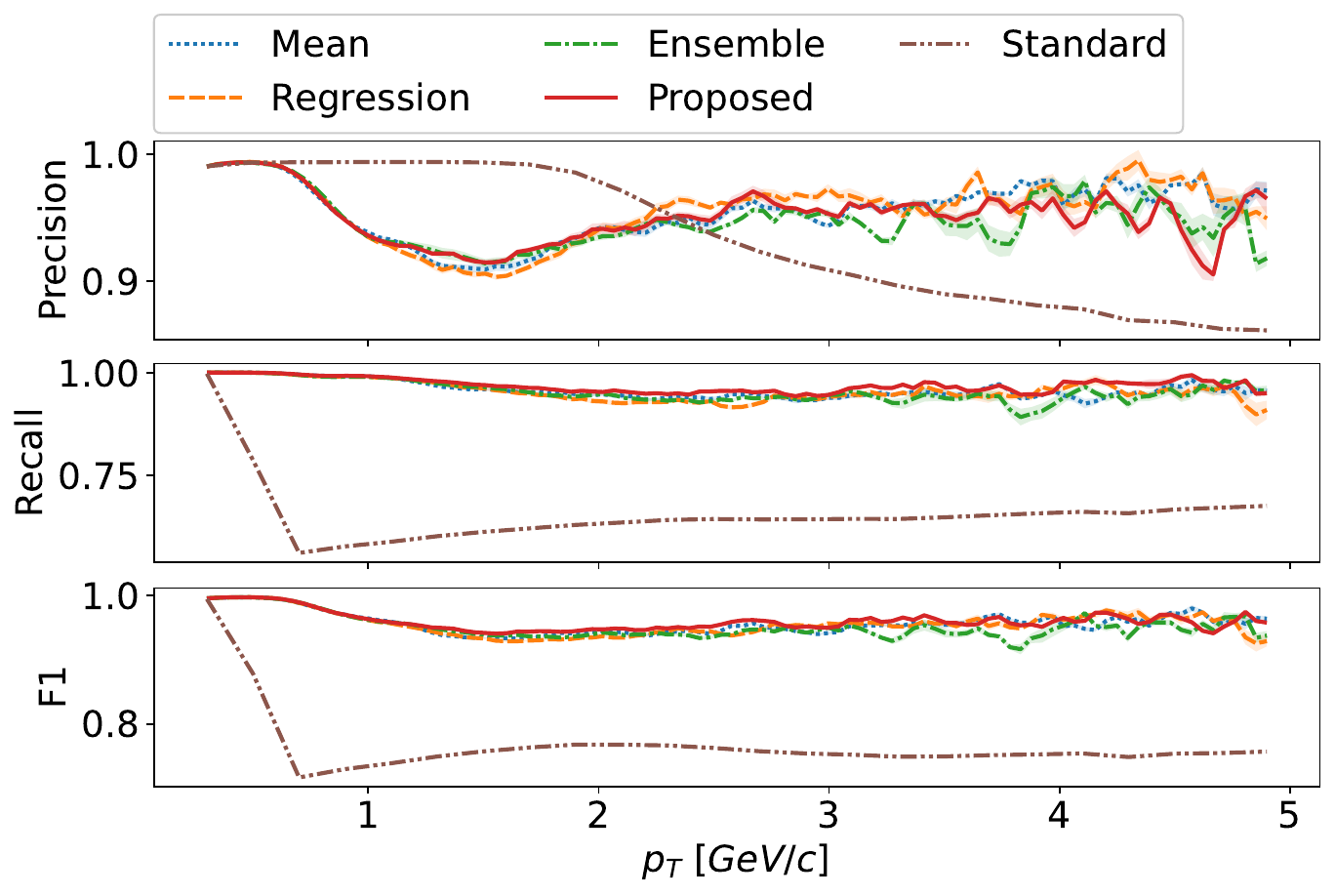}
      \captionof{figure}{Performance of different PID methods in pion selection task with missing data as a function of particle momentum.}
      \label{fig:pt_kaon_all}
    \end{minipage}
    \hfill
    \begin{minipage}[t]{.37\textwidth}
      \centering
      \includegraphics[width=\linewidth]{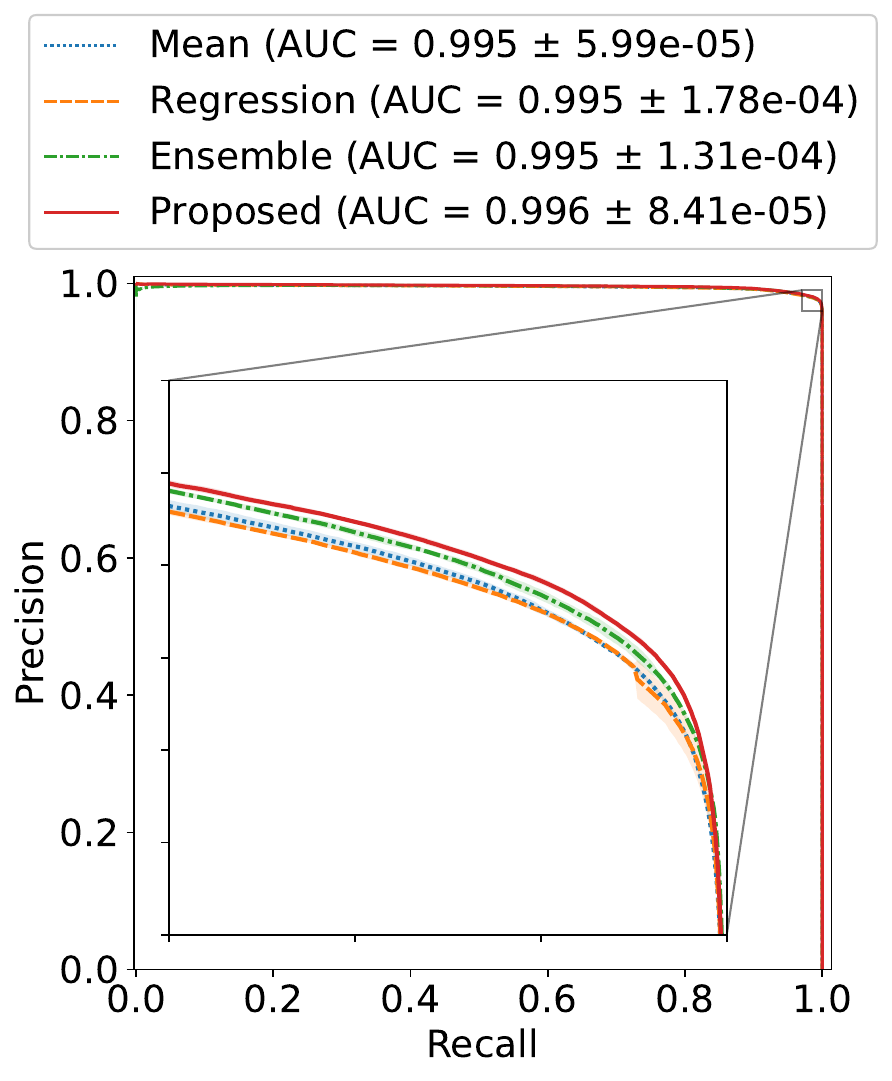}
      \captionof{figure}{Precision recall curve for different ML based approaches with missing data}
      \label{fig:prc_kaon_all}
    \end{minipage}
\end{figure}

\begin{figure}[h!]
    \begin{minipage}[t]{.58\textwidth}
      \centering
      \includegraphics[width=\linewidth]{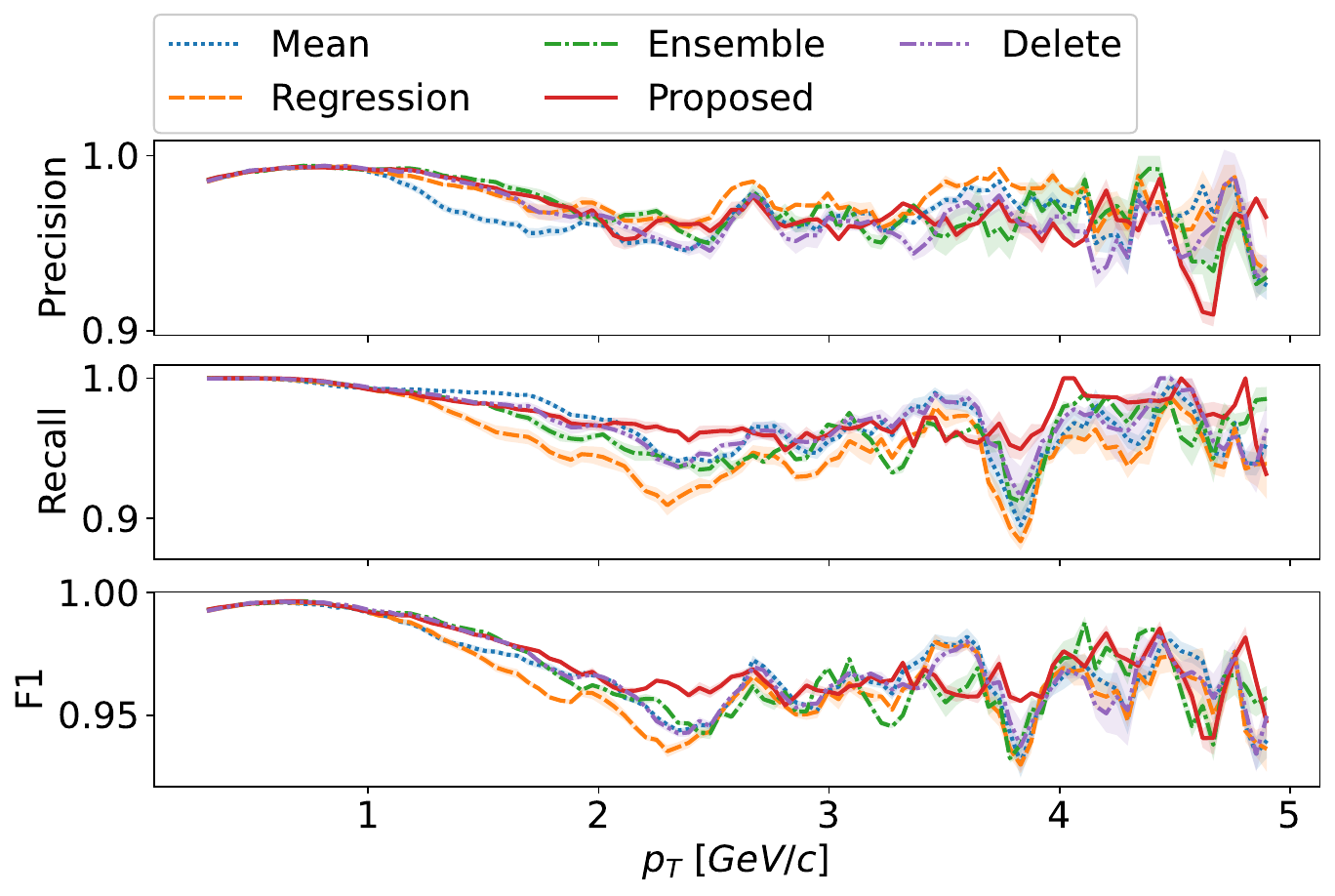}
      \captionof{figure}{Performance of different PID methods in pion selection task without missing data as a function of particle momentum.}
      \label{fig:pt_kaon_complete}
    \end{minipage}
    \hfill
    \begin{minipage}[t]{.37\textwidth}
      \centering
      \includegraphics[width=\linewidth]{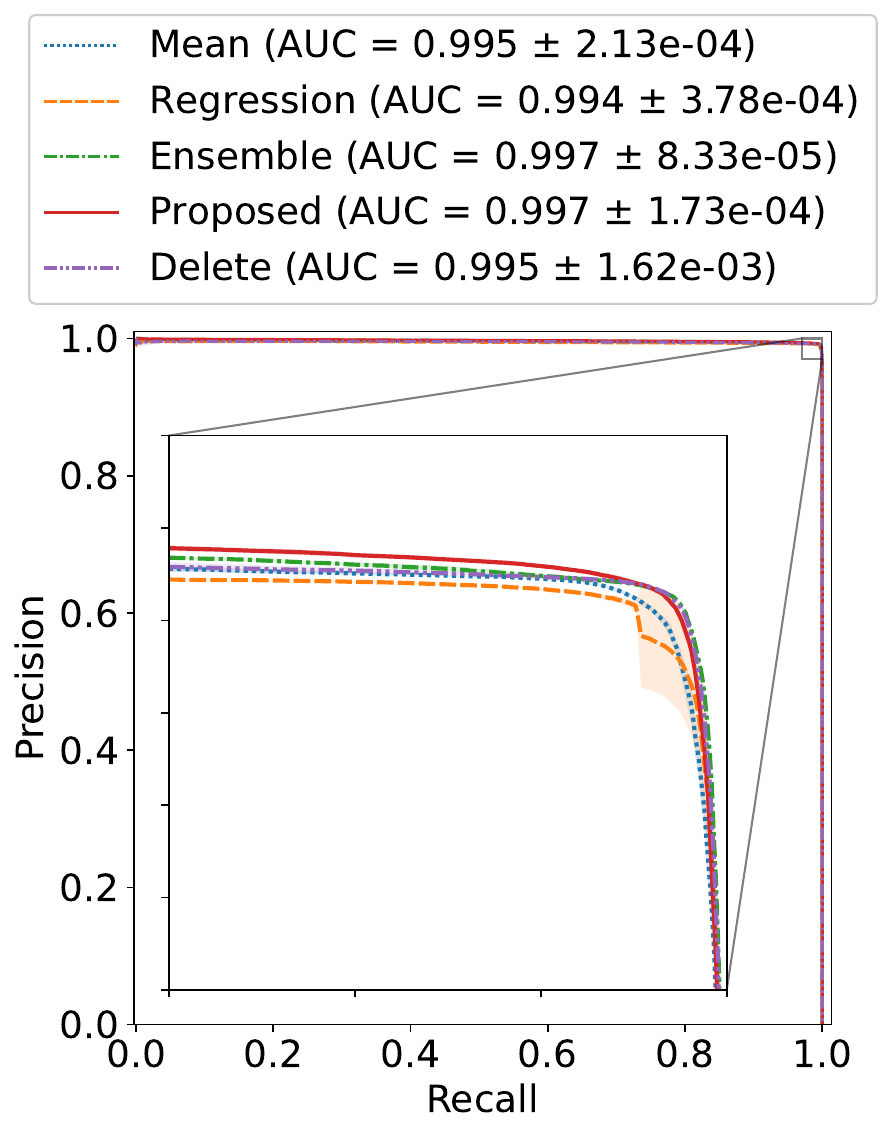}
      \captionof{figure}{Precision recall curve for different ML based approaches without missing data}
      \label{fig:prc_kaon_complete}
    \end{minipage}
\end{figure}

\newpage
\subsection{Antikaons}

\begin{figure}[h!]
    \begin{minipage}[t]{.58\textwidth}
      \centering
      \includegraphics[width=\linewidth]{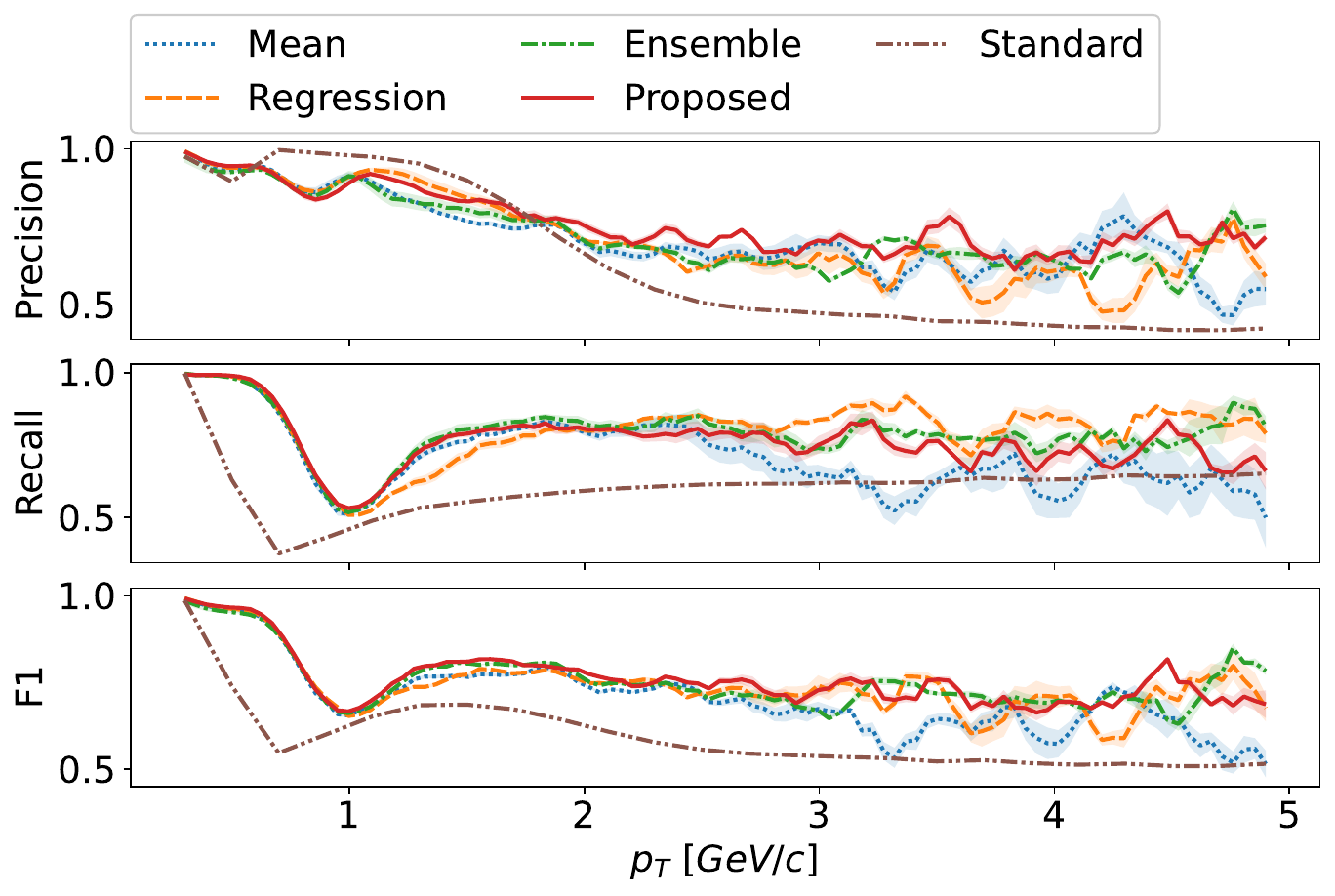}
      \captionof{figure}{Performance of different PID methods in antikaon selection task with missing data as a function of particle momentum.}
      \label{fig:pt_kaon_all}
    \end{minipage}
    \hfill
    \begin{minipage}[t]{.37\textwidth}
      \centering
      \includegraphics[width=\linewidth]{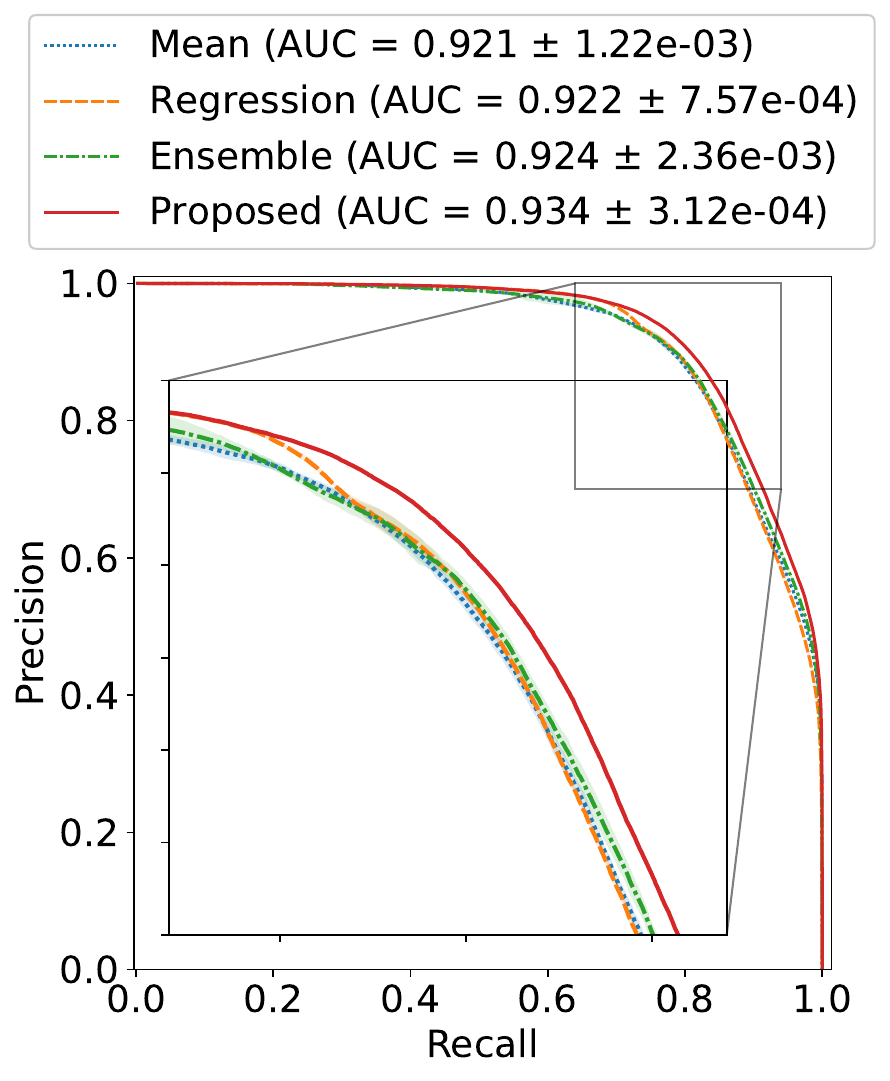}
      \captionof{figure}{Precision recall curve for different ML based approaches with missing data}
      \label{fig:prc_kaon_all}
    \end{minipage}
\end{figure}

\begin{figure}[h!]
    \begin{minipage}[t]{.58\textwidth}
      \centering
      \includegraphics[width=\linewidth]{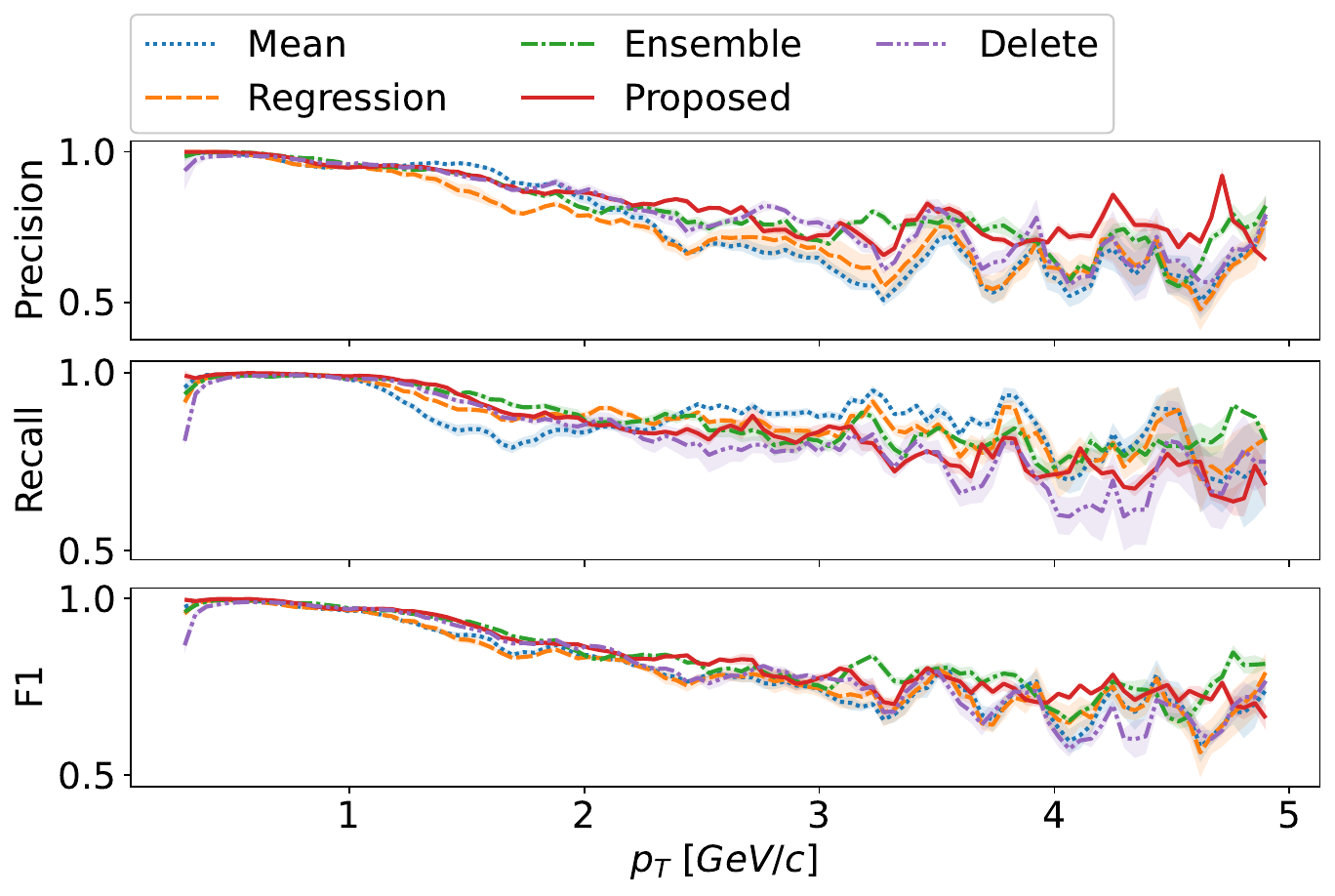}
      \captionof{figure}{Performance of different PID methods in antikaon selection task without missing data as a function of particle momentum.}
      \label{fig:pt_kaon_complete}
    \end{minipage}
    \hfill
    \begin{minipage}[t]{.37\textwidth}
      \centering
      \includegraphics[width=\linewidth]{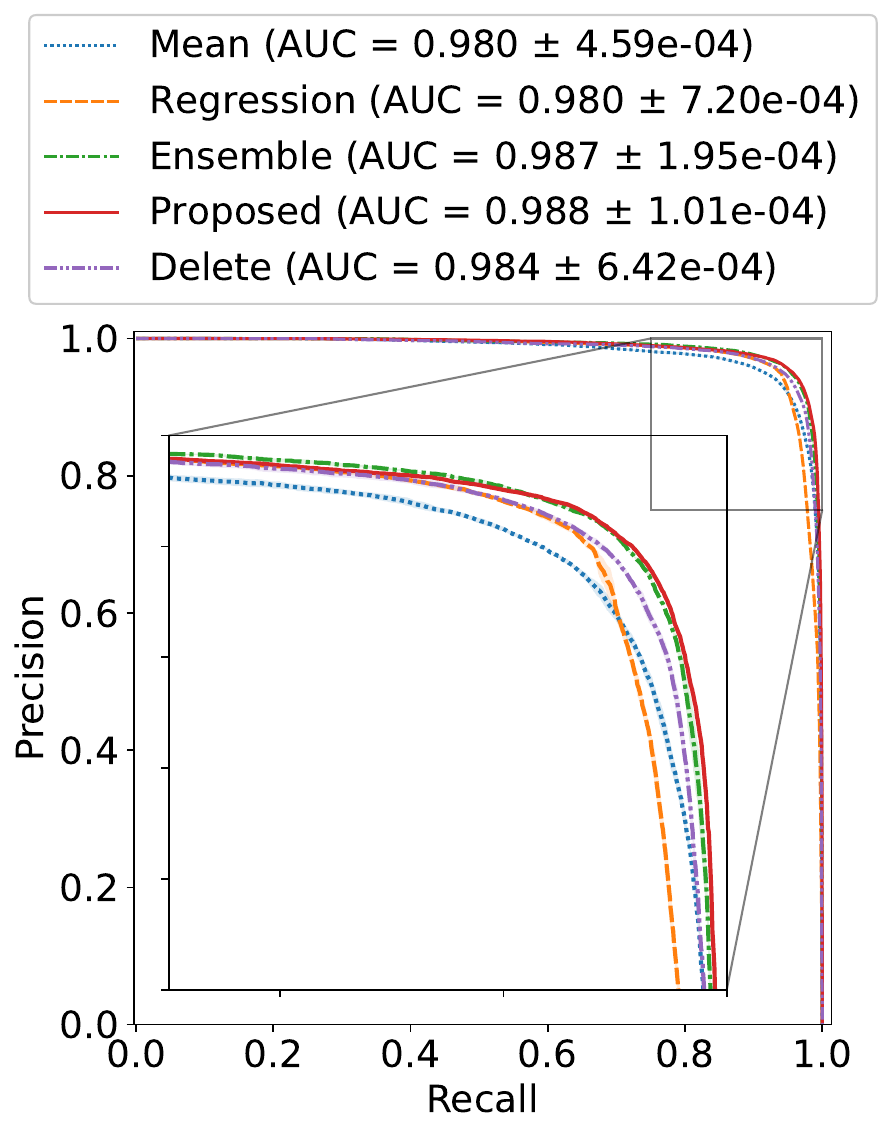}
      \captionof{figure}{Precision recall curve for different ML based approaches without missing data}
      \label{fig:prc_kaon_complete}
    \end{minipage}
\end{figure}

\newpage
\subsection{Antiprotons}

\begin{figure}[h!]
    \begin{minipage}[t]{.58\textwidth}
      \centering
      \includegraphics[width=\linewidth]{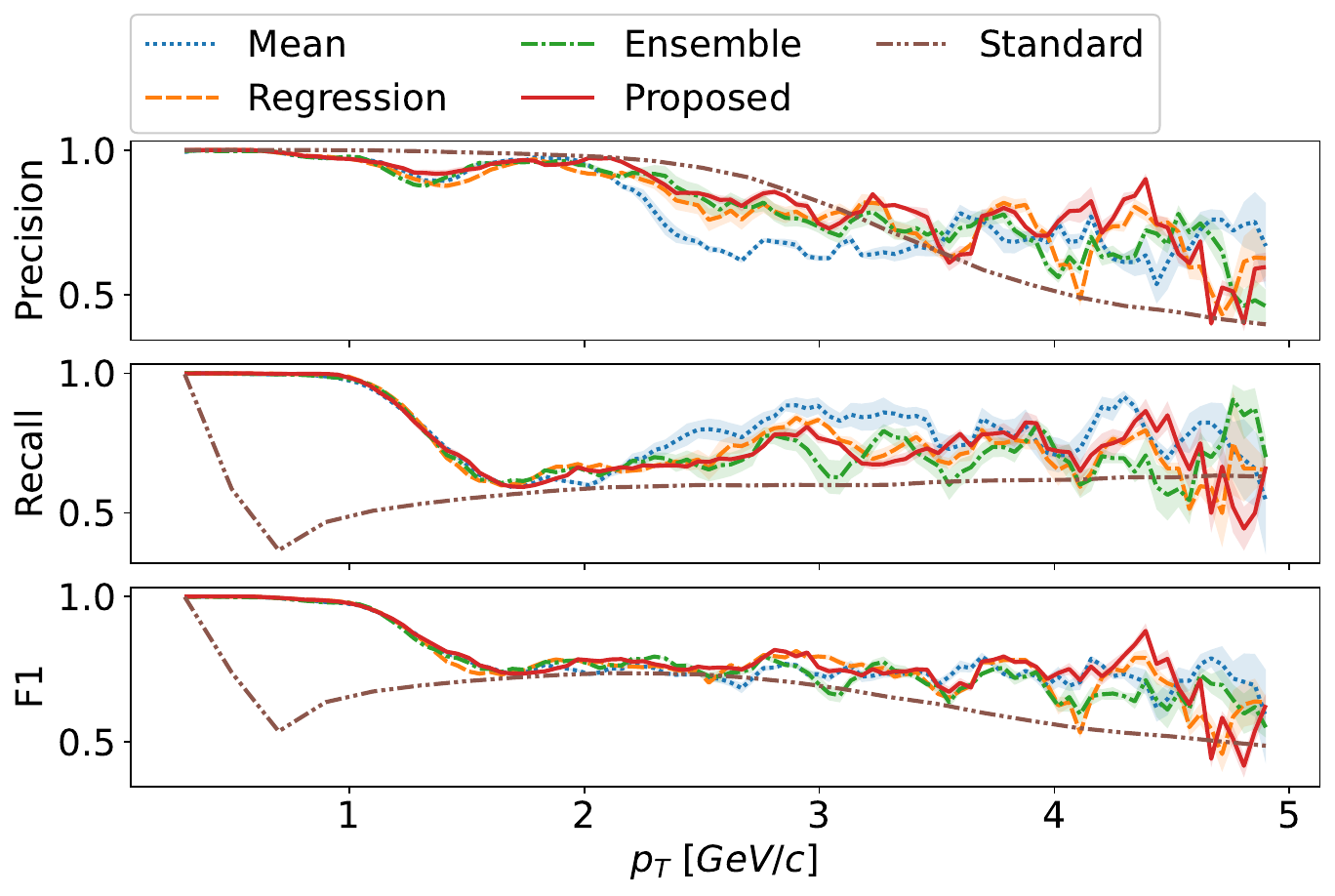}
      \captionof{figure}{Performance of different PID methods in antiproton selection task with missing data as a function of particle momentum.}
      \label{fig:pt_kaon_all}
    \end{minipage}
    \hfill
    \begin{minipage}[t]{.37\textwidth}
      \centering
      \includegraphics[width=\linewidth]{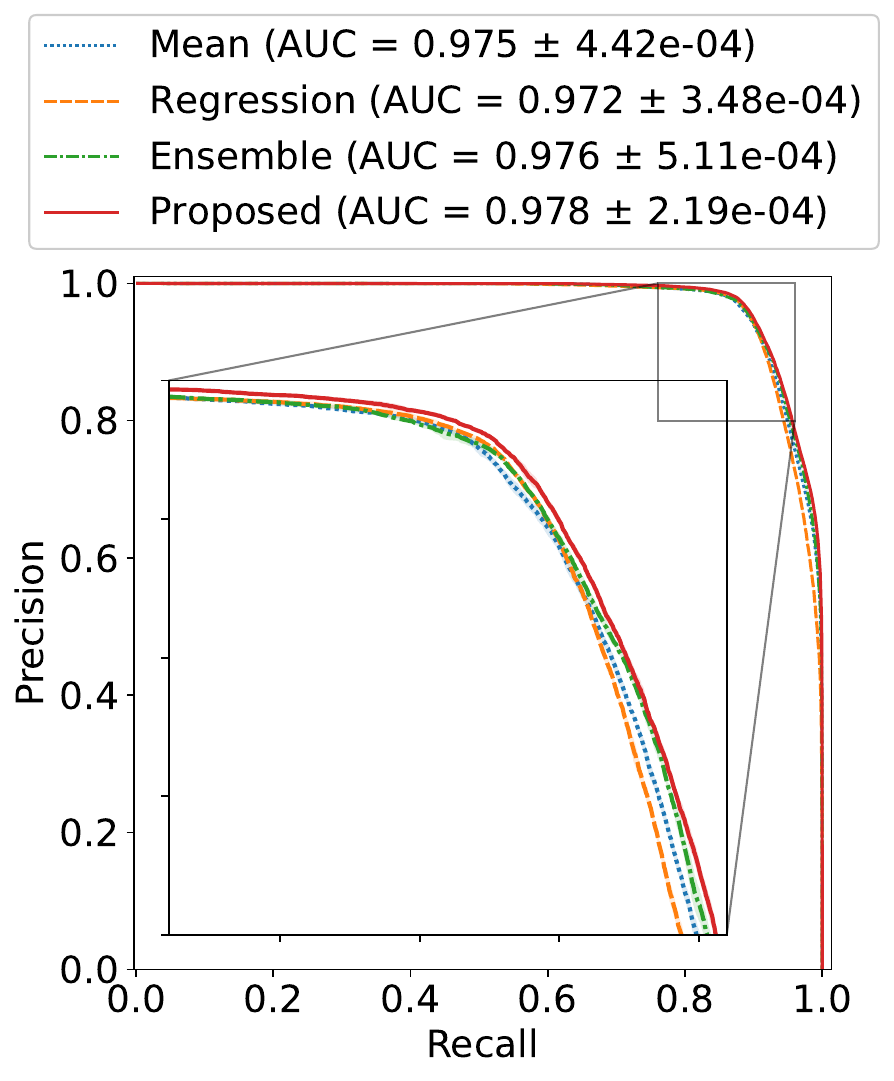}
      \captionof{figure}{Precision recall curve for different ML based approaches with missing data}
      \label{fig:prc_kaon_all}
    \end{minipage}
\end{figure}

\begin{figure}[h!]
    \begin{minipage}[t]{.58\textwidth}
      \centering
      \includegraphics[width=\linewidth]{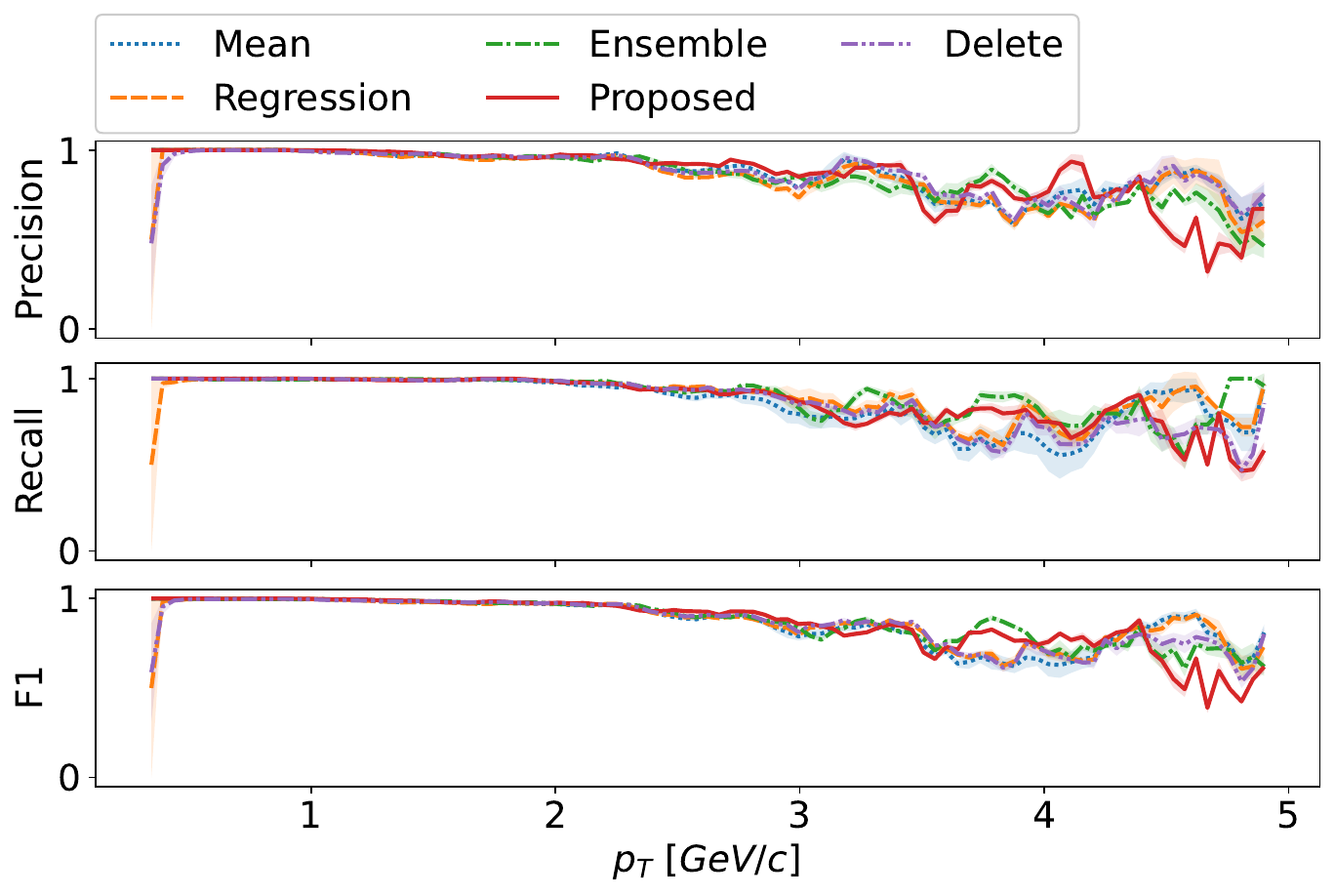}
      \captionof{figure}{Performance of different PID methods in antiproton selection task without missing data as a function of particle momentum.}
      \label{fig:pt_kaon_complete}
    \end{minipage}
    \hfill
    \begin{minipage}[t]{.37\textwidth}
      \centering
      \includegraphics[width=\linewidth]{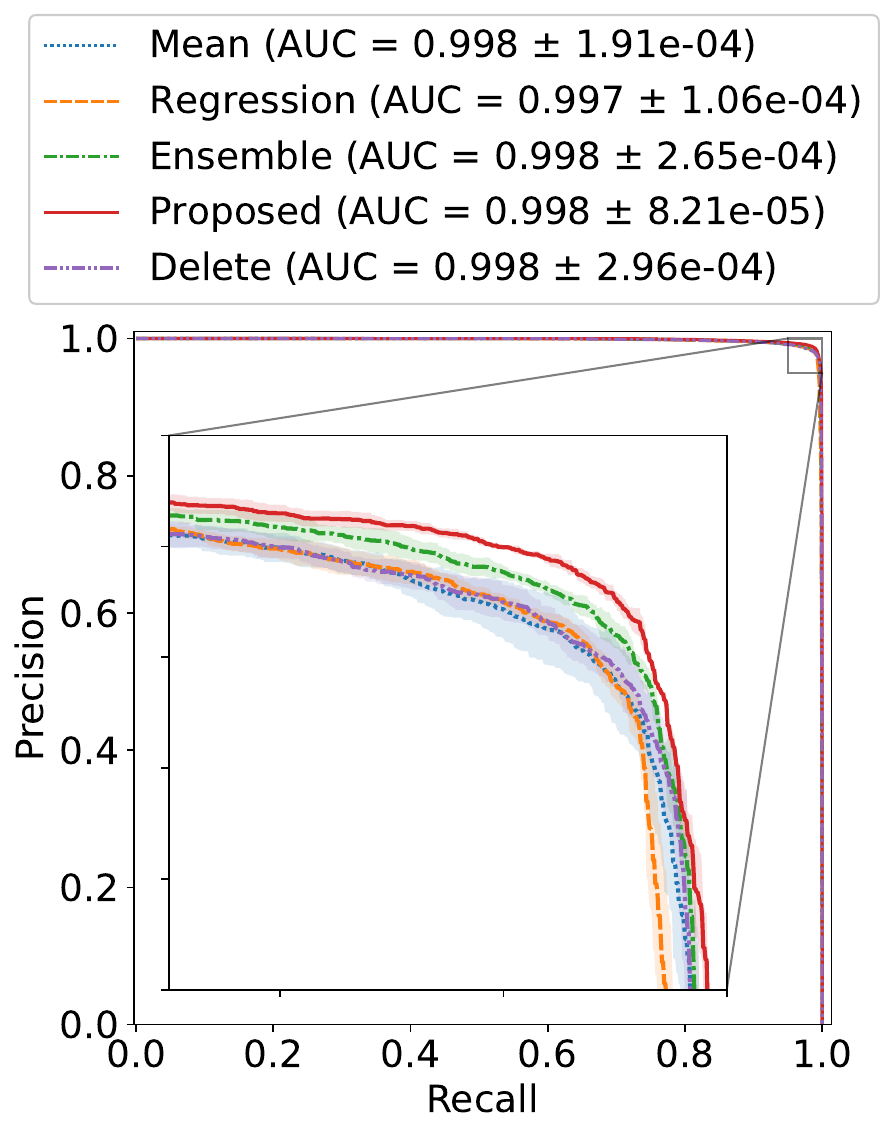}
      \captionof{figure}{Precision recall curve for different ML based approaches without missing data}
      \label{fig:prc_kaon_complete}
    \end{minipage}
\end{figure}

\newpage
\subsection{Antipions}

\begin{figure}[h!]
    \begin{minipage}[t]{.58\textwidth}
      \centering
      \includegraphics[width=\linewidth]{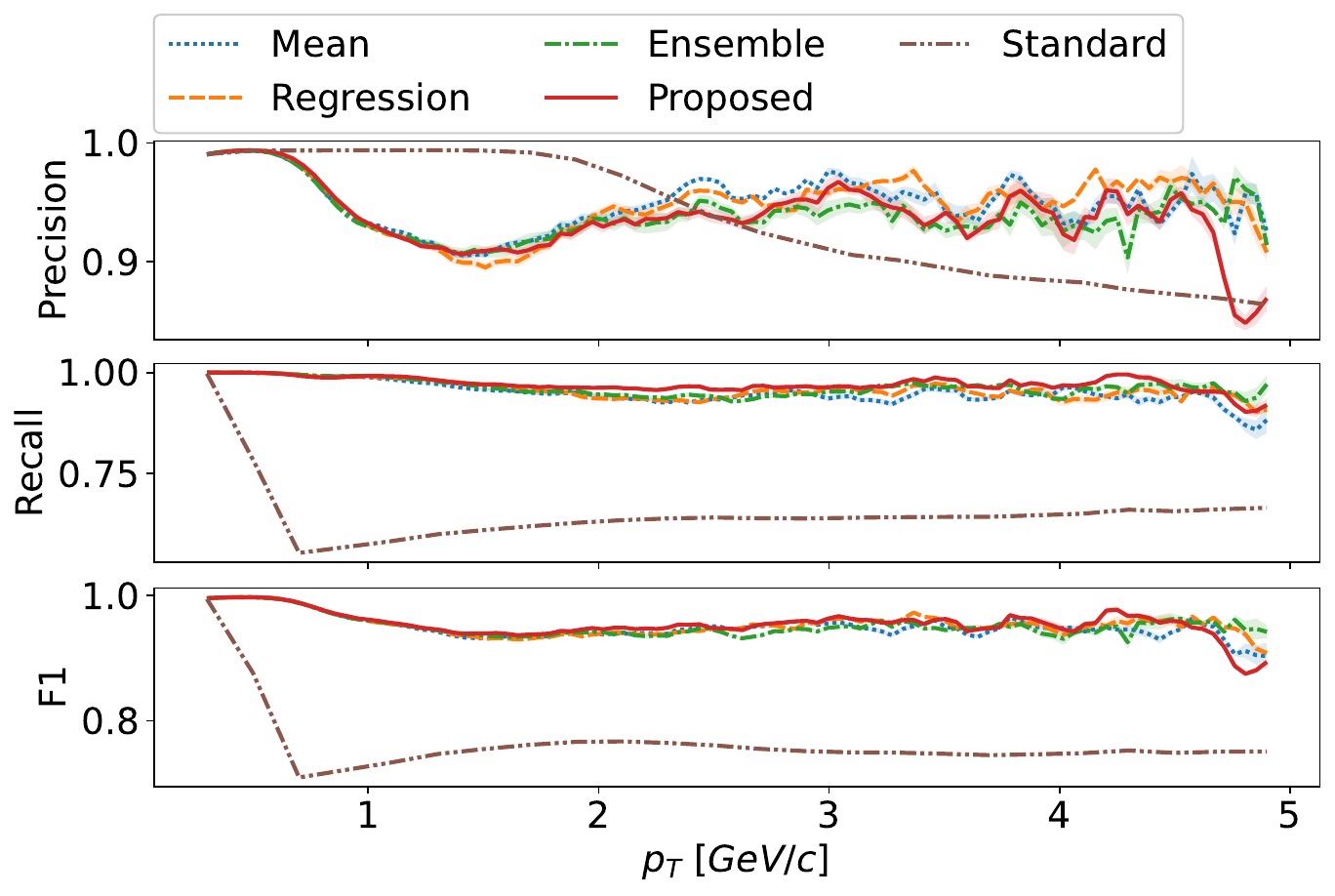}
      \captionof{figure}{Performance of different PID methods in antipion selection task with missing data as a function of particle momentum.}
      \label{fig:pt_kaon_all}
    \end{minipage}
    \hfill
    \begin{minipage}[t]{.37\textwidth}
      \centering
      \includegraphics[width=\linewidth]{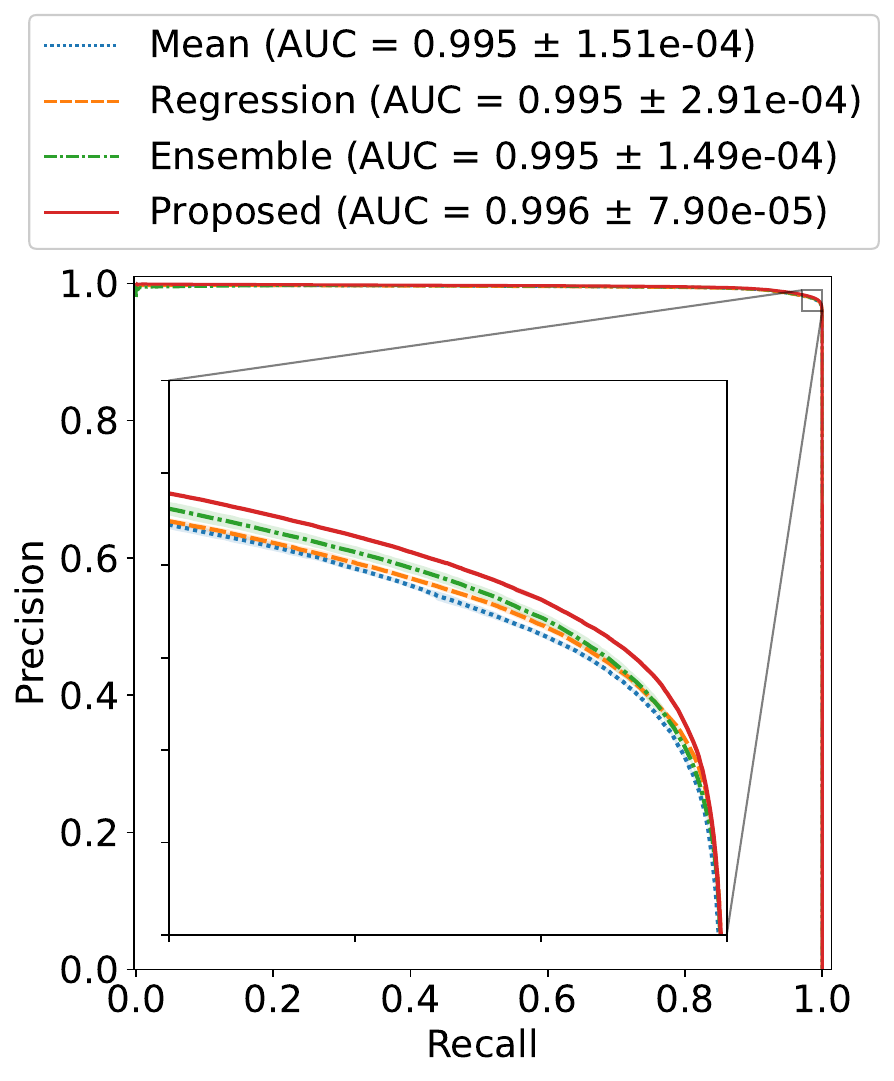}
      \captionof{figure}{Precision recall curve for different ML based approaches with missing data}
      \label{fig:prc_kaon_all}
    \end{minipage}
\end{figure}

\begin{figure}[h!]
    \begin{minipage}[t]{.58\textwidth}
      \centering
      \includegraphics[width=\linewidth]{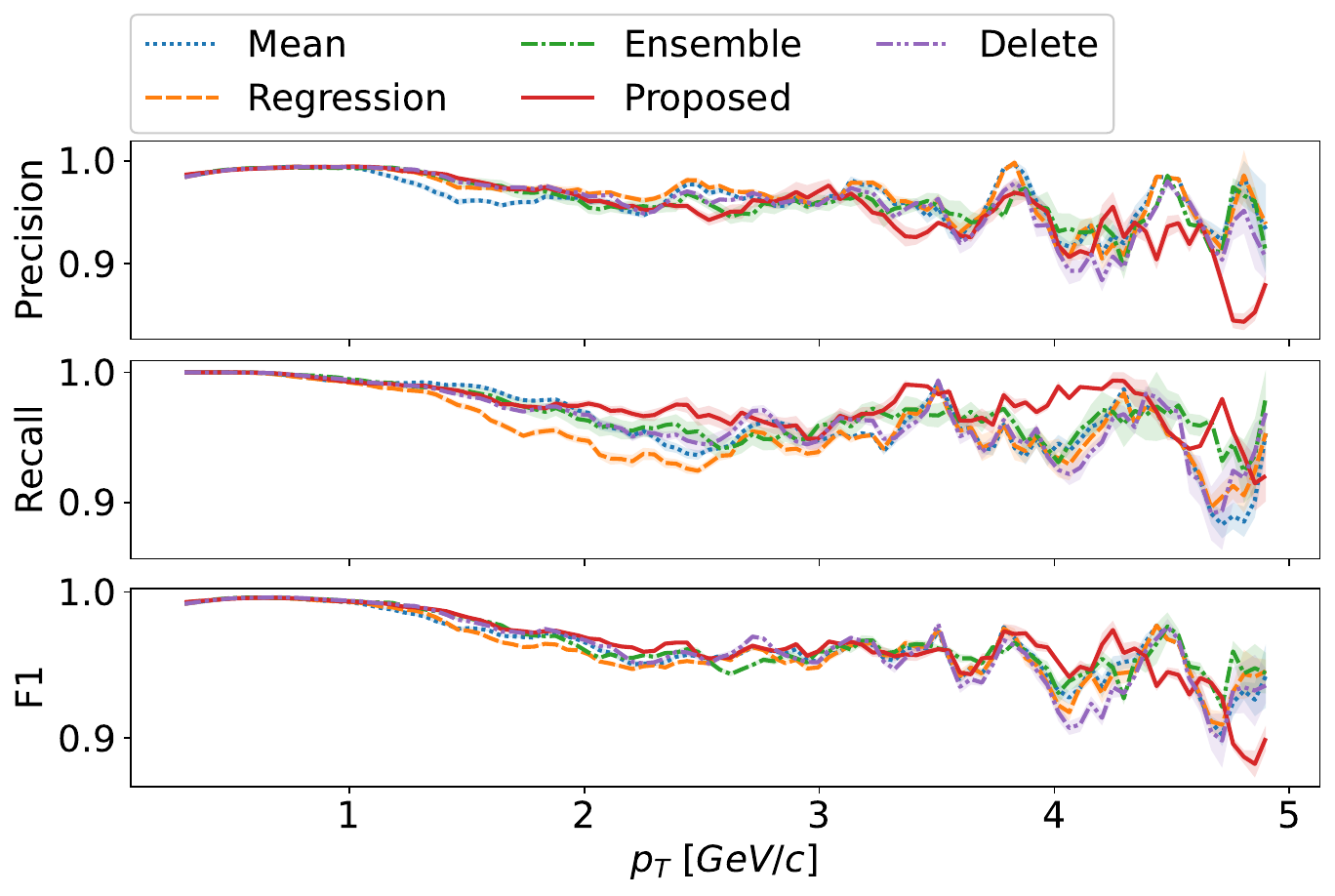}
      \captionof{figure}{Performance of different PID methods in antipion selection task without missing data as a function of particle momentum.}
      \label{fig:pt_kaon_complete}
    \end{minipage}
    \hfill
    \begin{minipage}[t]{.37\textwidth}
      \centering
      \includegraphics[width=\linewidth]{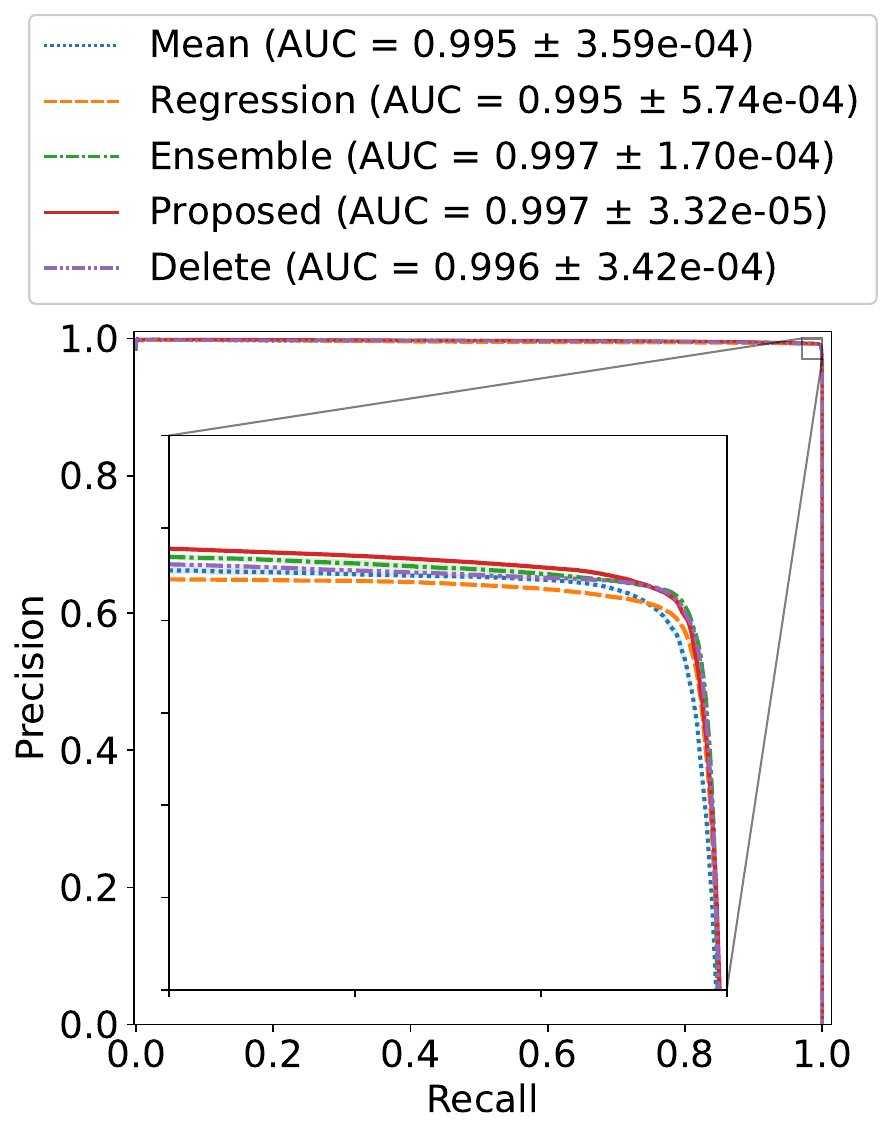}
      \captionof{figure}{Precision recall curve for different ML based approaches without missing data}
      \label{fig:prc_kaon_complete}
    \end{minipage}
\end{figure}